\documentclass[11pt]{article}

\usepackage[margin=1in]{geometry}
\usepackage{setspace}
\usepackage{graphicx}
\usepackage{soul}
\usepackage{comment}
\usepackage{booktabs}
\usepackage{makecell}
\usepackage{tabularx}
\usepackage{array}
\usepackage{mathtools}
\usepackage{xcolor}
\usepackage{url}
\usepackage{braket}
\usepackage{caption}

\usepackage{newtxtext,newtxmath}

\usepackage[numbers,sort&compress]{natbib}
\setcitestyle{super}

\newcommand{\natphysbibstyle}{unsrtnat}
\IfFileExists{naturemag.bst}{\renewcommand{\natphysbibstyle}{naturemag}}{}
\IfFileExists{sn-mathphys-num.bst}{\renewcommand{\natphysbibstyle}{sn-mathphys-num}}{}

\renewenvironment{abstract}
	{\quotation}
	{\endquotation}

\date{}

\makeatletter
\renewcommand{\fnum@table}{\textbf{Table \thetable}}
\makeatother
\DeclareCaptionLabelSeparator{npbar}{\space\ensuremath{\vert}\space}
\newif\ifincludeSI
\newif\ifonlySI

\includeSItrue
\onlySIfalse

\newif\ifshowSI
\ifincludeSI\showSItrue\fi
\ifonlySI\showSItrue\fi

\def\scititle{
    Engineering energy-time entanglement \\ from resonance fluorescence
}
\title{\bfseries \scititle}

\author{
    Jian~Wang$^{1\dagger}$,
	Xiu-Bin Liu$^{1,2\dagger}$,
    Ziqi Zeng$^{1,3,4}$,
	Xu-Jie Wang$^{1}$,
    Carlos Antón-Solanas$^{5,6}$ \\
    Li Liu$^{1}$,
    Hanqing~Liu$^{7,8}$,   
    Haiqiao~Ni$^{7,8}$, 
    Zhichuan~Niu$^{7,8}$,
    Bang Wu$^{1}$,
    Zhiliang~Yuan$^{1\ast}$\and
\small$^{1}$Beijing Academy of Quantum Information Sciences, Beijing 100193, China.\and
\small$^{2}$School of Physics and Beijing Key Laboratory of Opto-electronic Functional Materials and Micro-nano Devices,\and \small Key Laboratory of Quantum State Construction and Manipulation (Ministry of Education), \and \small Renmin University of China, Beijing 100872, China.\and
\small$^{3}$Beijing National Laboratory for Condensed Matter Physics, Institute of Physics, \and \small Chinese Academy of Sciences, Beijing 100190, China.\and
\small$^{4}$University of Chinese Academy of Sciences, Beijing 100049, China.\and
\small$^{5}$Departamento de Física de Materiales, Instituto Nicolás Cabrera,\and \small  Universidad Autónoma de Madrid, 28049 Madrid, Spain.\and
\small$^{6}$Condensed Matter Physics Center (IFIMAC), Universidad Autónoma de Madrid, Madrid 28049, Spain.\and
\small$^{7}$State Key Laboratory of Optoelectronic Materials and Devices, Institute of Semiconductors,\and \small Chinese Academy of Sciences,Beijing 100083, China.\and \small$^{8}$Center of Materials Science and Optoelectronics Engineering, \and \small University of Chinese Academy of Sciences, Beijing 100049, China.\and
\small$^{\ast}$Corresponding author. Email: yuanzl@baqis.ac.cn\\
\small$^{\dagger}$These authors contributed equally to this work.
}

\begin{document} 
%\linenumbers
\ifonlySI
\else
% Insert the title and author list
\maketitle

% Abstract, in bold
% There are strict length limits, and not all formats have abstracts.
% Consult the journal instructions to authors for details.
% Do not cite any references in the abstract.

\begin{abstract} \bfseries \boldmath

Resonance fluorescence from a coherently driven two-level emitter is a minimal quantum optical field that combines phase coherence with single-photon-level nonlinearity. Here we show that it can be engineered, using only passive linear interferometry, into energy-time entanglement. By injecting resonance fluorescence from a single quantum dot into an asymmetric Mach--Zehnder interferometer operated near destructive interference of the single-photon component, we generate an output field whose coincidence statistics are dominated by the simultaneous two-photon contribution $\ket{2}$ and the temporally separated photon-pair contribution $\ket{11}$. In a Franson geometry, these two sectors are resolved on the coincidence-delay axis, and both exhibit high-visibility nonlocal interference fringes and violate the Clauser--Horne--Shimony--Holt Bell inequality. Our results reveal a general route for engineering entanglement from resonance fluorescence using passive optics.

\end{abstract}

% Opening paragraph
% Nor is it indented
\noindent
Quantum entanglement is a foundational resource of quantum information science~\cite{Bennett2000,%Horodecki2009,
wehner2018quantum}. Since Franson's seminal proposal~\cite{franson1989bell}, time--energy entanglement has become a workhorse for quantum communications  
owing to its robustness against dispersion, compatibility with multiplexing and high-dimensional encoding, and ability to support device-independent protocols~\cite{xavier2025energy,zhong2024hyperentanglement,%yu2025quantum,
singh2025photonic}.
Traditionally, entangled-photon sources have relied on spontaneous parametric processes~\cite{ShihAlley1988_PRL,OuMandel1988_PRL,Kwiat1995,Li2005FiberEntangled}. 
These sources have been extraordinarily successful and remain central to quantum optics and photonic quantum information processing~\cite{Carolan2015,Yin2017,zhong2020quantum,Bao2023VLSI, Bourassa2025PsiQuantum}, despite their probabilistic nature and the superlinear growth of multi-pair contributions with brightness.

Complementary to parametric sources, a conceptually distinct route is offered by resonance fluorescence (RF) from a two-level emitter~\cite{peiris2017franson}.
Recent advances have shown that this elementary platform can generate photon pairs that violate Bell inequalities in the context of two-photon scattering~\cite{liu2024violation,wang2025purcell}, alongside a broader range of rich multiphoton phenomena~\cite{LopezCarreno2018JointSubnatural,masters2023simultaneous,wang2025coherence, kim2025unlocking,bracht2025tunable}.
These developments are closely tied to renewed interest in the weak-excitation (Heitler) regime~\cite{LopezCarreno2018JointSubnatural,phillips2020photon,hanschke2020origin}. In this limit, the emission is strongly antibunched on the timescale of the radiative lifetime, yet remains phase-referenced to the excitation laser and inherits its linewidth~\cite{hoffges1997heterodyne,Nguyen2011UltraCoherentSPS,matthiesen2013phase}. 
This confluence of properties motivates a pure-state description~\cite{wang2025coherence} in which the driven emitter and the emitted field are treated jointly as a pure state. 
In this picture, laser-like linewidth and photon antibunching emerge without invoking higher-order scattering processes~\cite{wang2025coherence,HuangGQ2026}.
The same perspective is also relevant to interferometric schemes that reshape the photon-number statistics of RF, including our self-homodyne approach~\cite{wang2025coherence} and other experiments based on interference between RF and a local oscillator~\cite{kim2025unlocking,bracht2025tunable,Yin2026TunableHighOrderCoherence}. More broadly, it suggests that passive interferometric networks can transform  Heitler-regime RF into nonclassical multiphoton states, and open a route toward genuine multiphoton entanglement, beyond beam-splitter-based single-photon entanglement schemes~\cite{hu2025transforming,Zeng2025Bell}.

In this work, we demonstrate linear-optical synthesis of time--energy entangled states from the RF of a single quantum dot (QD) using solely an asymmetric Mach--Zehnder interferometer (AMZI).
We test the resulting states in a Franson interferometer and observe high-visibility two-photon interference together with a clear violation of the Clauser--Horne--Shimony--Holt (CHSH) Bell inequality. Our results show that the simplest form of RF, when combined with passive linear optics, provides a conceptually distinct and resource-efficient route to robust entanglement generation across emitter platforms, from quantum dots and atoms.

%%%%%%%%%%%%%%%% MAIN TEXT FIGURES %%%%%%%%%%%%%%%
\begin{figure}
\centering 
\includegraphics[width=0.9\textwidth]{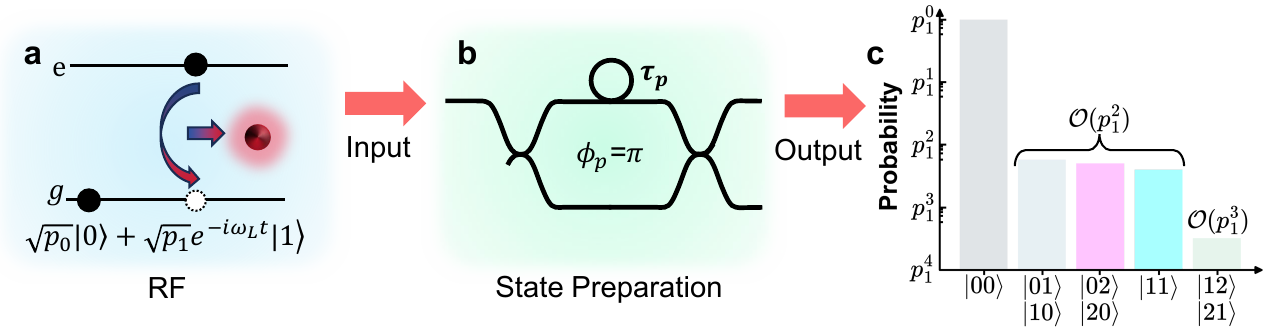}
\caption{\textbf{Interferometric synthesis of time--energy entanglement.}
The RF field from a coherently driven two-level emitter is injected into an asymmetric Mach--Zehnder interferometer (AMZI) with phase set to $\pi$.
Destructive interference suppresses the probability of single-photon terms from $\mathcal{O}(p_1)$ to $\mathcal{O}(p_1^2)$, reshaping the output toward prominence of two-photon components: a same-bin pair $\ket{2}$ ($\ket{02}$ or $\ket{20}$) and a time-bin pair $\ket{11}$ separated by the interferometer delay imbalance $\tau_p$.
Because the emission time of each RF photon is indeterminate prior to detection, the temporal superposition of these two-photon amplitudes yields time--energy entanglement. $p_0$ and $p_1$ are the probabilities of the vacuum and one-photon components, respectively, satisfying $p_0 + p_1 = 1$.}
\label{fig1}
\end{figure}

\section*{Preparation of distinct two-photon states}
Figure~\ref{fig1}\textbf{a} illustrates the RF from a two-level emitter. In the pure-state model~\cite{wang2025coherence}, each temporal mode of the RF field is represented as a vacuum--single-photon superposition, $\ket{\psi} = \sqrt{p_0}\ket{0} + \sqrt{p_1}\ket{1}$, phase-locked to the driving laser.
Although the full description involves an entangled emitter-field state, this effective representation captures the field's essential coherence and provides an accurate, simplified basis for deriving the first-order coherence and multi-photon correlations~\cite{wang2025coherence,HuangGQ2026,Yin2026TunableHighOrderCoherence}. When injected into an AMZI (Fig.~\ref{fig1}\textbf{b}), emission amplitudes separated by the interferometer delay imbalance ($\tau_p$) interfere to form photon-number states $\ket{n_t\, n_{t+\tau_p}}$ across two temporal modes (time bins), with the photon number in each mode truncated at $n=2$. Crucially, setting the interferometer phase to $\pi$ suppresses the single-photon components $\ket{01}$ and $\ket{10}$ from scaling linearly in $p_1$ to scaling quadratically in $p_1$, placing them on the same order as the two-photon terms [Fig.~\ref{fig1}\textbf{c}]. Consequently, the $\ket{2}$ and $\ket{11}$ components dominate the subsequent two-photon coincidence measurements. For brevity, we use $\ket{2}$ to denote the ``same-bin'' two-photon sector, encompassing both $\ket{02}$ and $\ket{20}$. 
Owing to the indeterminate nature of the emission time of each RF photon, the prepared $\ket{2}$ and $\ket{11}$ components are intrinsically time--energy entangled, analogous to photon pairs from a cw-pumped parametric down-conversion source~\cite{Rarity1990PRL}; however, the single-emitter platform does not generate multi-pair events.
The full theoretical derivation of the corresponding two-photon components is given in Supplementary Information 1.3.

\begin{figure}[hbt]
\centering 
\includegraphics[width=0.88\textwidth]{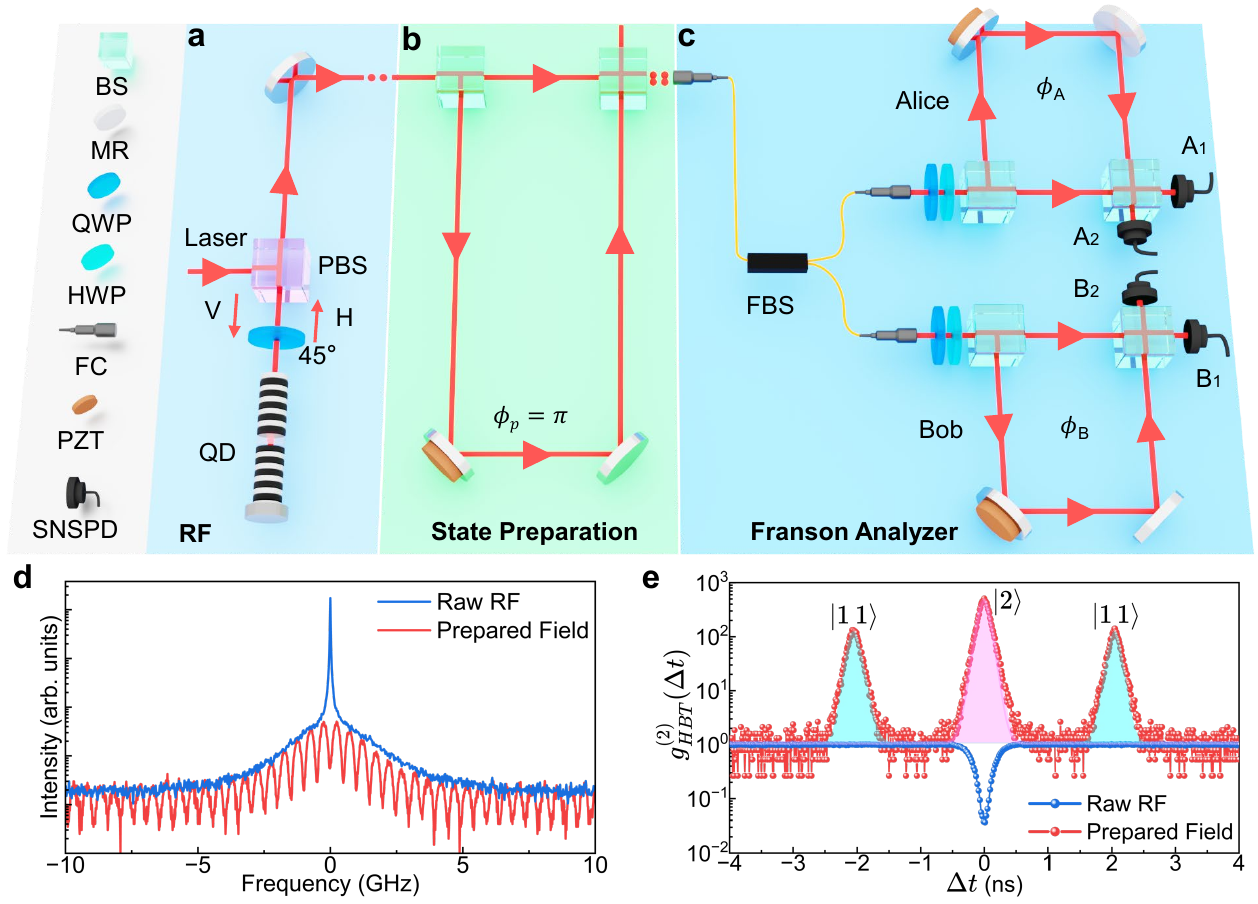}
\caption{\textbf{Experimental setup and state preparation.}
\textbf{a}, RF source: an InAs QD in a low-reflectivity micropillar cavity under cw resonant excitation.
\textbf{b}, State-preparation AMZI.
\textbf{c}, Franson analyser.
\textbf{d}, High-resolution spectra of the raw RF (blue) and the field after the state-preparation AMZI (red); traces are vertically offset for clarity.
\textbf{e}, Second-order autocorrelation functions $g^{(2)}_{HBT}(\Delta t)$ of the raw RF (blue) and the AMZI-prepared output (red).
Abbreviations:  FBS, fiber beam splitter; PBS, polarizing beam splitter; BS, beam splitter; MR, mirror; QWP/HWP, quarter-/half-wave plate; FC, fiber coupler; PZT, piezoelectric transducer; SNSPD, superconducting nanowire single-photon detector.}
\label{fig2}
\end{figure} 

Our experimental setup is summarized in Fig.~\ref{fig2}. The RF is generated from an InAs QD embedded in a low-reflectivity micropillar cavity under cw resonant excitation, following the scheme of Ref.~\cite{wu2023mollow}. The emitted RF is sent to the state-preparation AMZI with imbalance $\tau_p=2.14~\mathrm{ns}$ and phase set to $\pi$. The prepared field is then routed by a fibre beam splitter (FBS) to a Franson analyzer, where Alice and Bob each implement a matched analysis AMZI and record coincidences using two single-photon detectors. To prevent re-interference of the preparation time bins, we choose the analysis delay $\tau_m \approx \tau_p/2 = 1.07~\mathrm{ns}$.  Both $\tau_p$ and $\tau_m$ are chosen to be much longer than the QD coherence time ($\tau_p,\tau_m \gg T_2$) so that the action of all three interferometers does not perturb the emitter’s steady-state dynamics. Further details of the QD–micropillar device and the AMZI's phase locking are provided in Methods and in Supplementary Information 2.1, respectively.

In Fig.~\ref{fig2}\textbf{d}, we compare the high-resolution spectra of the raw RF (blue) and the prepared field after the state-preparation AMZI (red). 
These measurements were performed on a neutral exciton transition at $\nu=328.91~\mathrm{THz}$ with driving powers of mean photon numbers of $\bar{n}=0.05$ and $0.01$, respectively, where $\bar{n}=\frac{P_{\text{in}}T_1}{h\nu}$ is evaluated from the measured input power $P_{\text{in}}$ and the exciton lifetime $T_1=67.2~\mathrm{ps}$. The raw RF spectrum, plotted versus frequency detuning from $\nu$, shows a narrow, laser-like peak superimposed on a broader background, often discussed in the literature in terms of `coherent' and `incoherent' contributions~\cite{Mollow1969}.
After the state-preparation AMZI, the laser-like component is strongly suppressed by destructive interference, leaving a broadband spectrum modulated by interference fringes whose spacing ($\sim$467~MHz) corresponds to the AMZI imbalance  ($\tau_p\simeq2.14$~ns). A detailed dependence of the filtered spectrum on the AMZI phase $\phi_p$ and excitation strength is provided in Supplementary Information 2.2.

The state-preparation AMZI leads to a pronounced change in the photon-number statistics of the RF, as illustrated by a Hanbury Brown--Twiss (HBT) measurement at an excitation flux of $\bar{n}=0.01$. As shown in Fig.~\ref{fig2}\textbf{e}, the raw RF (blue) is strongly antibunched at zero delay, whereas the AMZI-prepared field (red) displays a striking super-bunching at $\Delta t=0$ with $g^{(2)}_{HBT}(0)\approx 511$, alongside satellite peaks at $\Delta t=\pm\tau_p$ where $g^{(2)}_{HBT}(\pm\tau_p)\approx 136$.
The side peaks arise from coincidences between photons occupying adjacent time bins separated by the state-preparation AMZI delay. 
These features show that the AMZI deterministically reshapes the two-photon correlations of the RF field.

\begin{figure}[hbt]
\centering 
\includegraphics[width=1\textwidth]{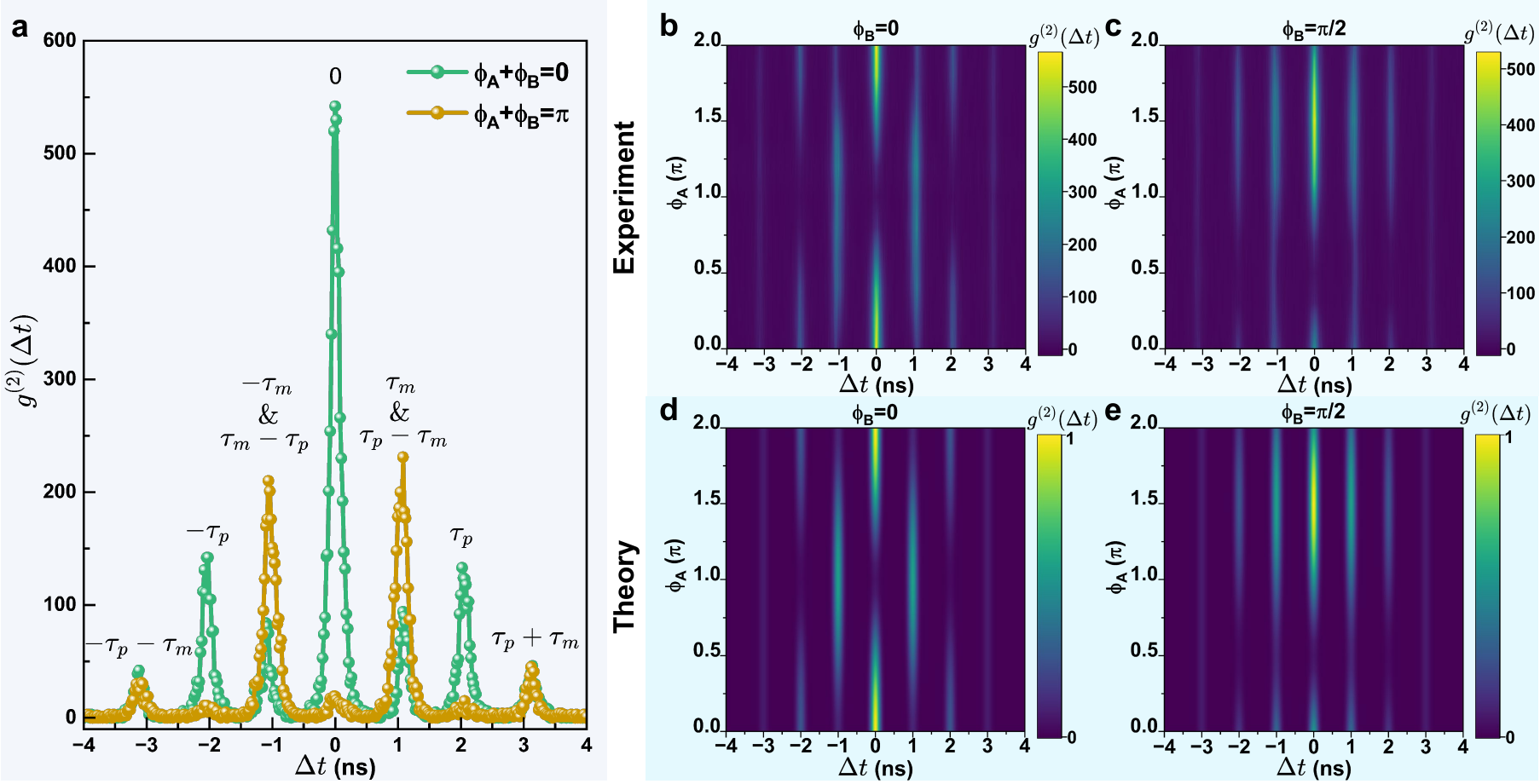}
\caption{\textbf{Franson interference.}
\textbf{a}, Measured second-order correlation functions between detectors $A_1$ and $B_1$ for phase settings $\phi_A + \phi_B = 0$ and $\pi$.
\textbf{b} and \textbf{c}, Measured two-dimensional correlation maps with $\phi_B = 0$ (\textbf{b}) and $\phi_B = \pi/2$ (\textbf{c}).
An excitation power of $\bar{n}=0.01$ was used for the measurements in \textbf{a--c}.
\textbf{d} and \textbf{e}, Corresponding theoretical simulations for \textbf{b} and \textbf{c}, respectively.}
\label{fig3}
\end{figure} 

\section*{Franson interference}

After the prepared field is injected into the Franson interferometer via the FBS (Fig.~\ref{fig2}\textbf{c}), each relevant two-photon component is coherently mapped onto the path-entangled Bell state 
\[
\frac{1}{\sqrt{2}}\left( \ket{s_A, s_B} + e^{i(\phi_A+\phi_B)} \ket{\ell_A, \ell_B} \right),
\]
%$\frac{1}{\sqrt{2}}\left( \ket{s_A, s_B} + e^{i(\phi_A+\phi_B)} \ket{\ell_A, \ell_B} \right)$, 
where $\phi_A$ ($\phi_B$) is the phase of Alice’s (Bob’s) analysis interferometer, and $\ket{s_A}$ ($\ket{\ell_A}$) and $\ket{s_B}$ ($\ket{\ell_B}$) denote propagation through the short (long) arms of the respective interferometers.
For clarity of presentation, we temporarily neglect the distinction between simultaneous and temporally separated photon pairs, and treat the $\ket{2}$ and $\ket{11}$ components equivalently as two-photon contributions in this Franson transformation.
To experimentally probe the coherence of these two-photon pathways, we independently tune the phases $\phi_{A}$ and $\phi_{B}$ at the Alice and Bob nodes to implement projective measurements on the time-entangled Bell states, and record the resulting coincidence counts. The $\ket{2}$ and $\ket{11}$ contributions manifest at distinct coincidence-delay peaks, allowing their two-photon interference signatures to be extracted separately.

Figure~\ref{fig3}\textbf{a} shows two representative coincidence histograms measured between detectors $A_1$ and $B_1$ for sum-phase settings $\phi_A+\phi_B=0$ and $\pi$. The data exhibit a characteristic multipeak structure, with pronounced features at $\Delta t=0$, $\pm\tau_m$, and $\pm\tau_p$, and weaker outer peaks at $\pm(\tau_p+\tau_m)$. Importantly, the strongest interference contrast occurs at $\Delta t=0$ and $\pm\tau_p$, where the coincidence rate switches between constructive and destructive interference as $\phi_A+\phi_B$ varies from $0$ to $\pi$, providing a clear signature of high-visibility two-photon interference. 
In comparison, the features near $\Delta t=\pm\tau_m$ arise from two timing contributions with unequal weights, as illustrated in Supplementary Fig.~S2: one at $\Delta t=\pm\tau_m$, and the other offset by the preparation-AMZI imbalance, yielding $\Delta t=\pm(\tau_p-\tau_m)$.
Because these two contributions overlap in the coincidence histogram but carry different amplitudes, so that their interference has noticeably poorer visibility.
The outermost peaks at $\pm(\tau_p+\tau_m)$ are essentially phase insensitive.

To map these phase responses systematically, we fix Bob’s phase at a given value and continuously scan Alice’s phase. Figures~\ref{fig3}\textbf{b} and \ref{fig3}\textbf{c} show the resulting two-dimensional correlation maps as a function of $\phi_A$ and $\Delta t$ for $\phi_B=0$ and $\pi/2$, respectively. Pronounced interference fringes are clearly visible at $\Delta t=0$ and $\pm\tau_p$, with the fringe phase governed by the sum phase $\phi_A+\phi_B$. 
In contrast, the fringes around $\Delta t=\pm\tau_m$ result from interference between the native $\Delta t=\pm\tau_m$ contributions and those associated with the $\pm(\tau_p-\tau_m)$ terms, and therefore shift in the opposite direction to those at $\Delta t=0$ and $\pm\tau_p$ for $\phi_B=0$.

The above observations can be understood by analysing the two-photon probability amplitudes contributing to each coincidence-delay peak. 
Using the pure-state model~\cite{wang2025coherence}, we derive the coincidence count probabilities at the characteristic delays, as detailed in Supplementary Information 1.4.
The central peak at $\Delta t=0$ and the peaks at $\Delta t=\pm\tau_p$ exhibit the strongest phase-dependent modulation governed by the sum phase $\phi_A+\phi_B$,
\begin{align}
\mathcal{C}(0) &\propto \frac{p_1^2}{128}\left[1+p_0^2\cos(\phi_A+\phi_B)\right], \label{Center}\\
\mathcal{C}(\pm \tau_p) &\propto \frac{p_1^2}{512}\left[1+p_0^2\cos(\phi_A+\phi_B)\right]. \label{2}
\end{align}
By contrast, the peaks around $\Delta t=\pm\tau_m$ display a weaker modulation that depends on the phase difference $\phi_A-\phi_B$,
\begin{equation}
\mathcal{C}(\pm \tau_m)\propto\frac{p_1^2}{1024}\left[ 5-4p_0^2\cos(\phi_A-\phi_B) \right],
\label{1}
\end{equation}
whereas the outer peaks at $\Delta t=\pm(\tau_m+\tau_p)$ remain phase independent,
\begin{equation}
\mathcal{C}\!\left[\pm(\tau_m+\tau_p)\right]\propto\frac{p_1^2}{1024}\left[ 2p_1(6p_1+1) +1\right].
\label{3}
\end{equation}
These relations directly account for the distinct fringe responses observed experimentally and highlight the coexistence of multiple two-photon pathways in the AMZI-engineered state. 
With measured photon-number statistics as inputs, our theory reproduces the phase-dependent coincidence maps in Figs.~\ref{fig3}\textbf{d} and \ref{fig3}\textbf{e}; the associated data-processing steps, including extraction of $p_1$ and concatenation of coincidence probabilities at different delays, are described in Supplementary Information~3.

\begin{figure}[hbt]
\centering 
\includegraphics[width=.7\columnwidth]{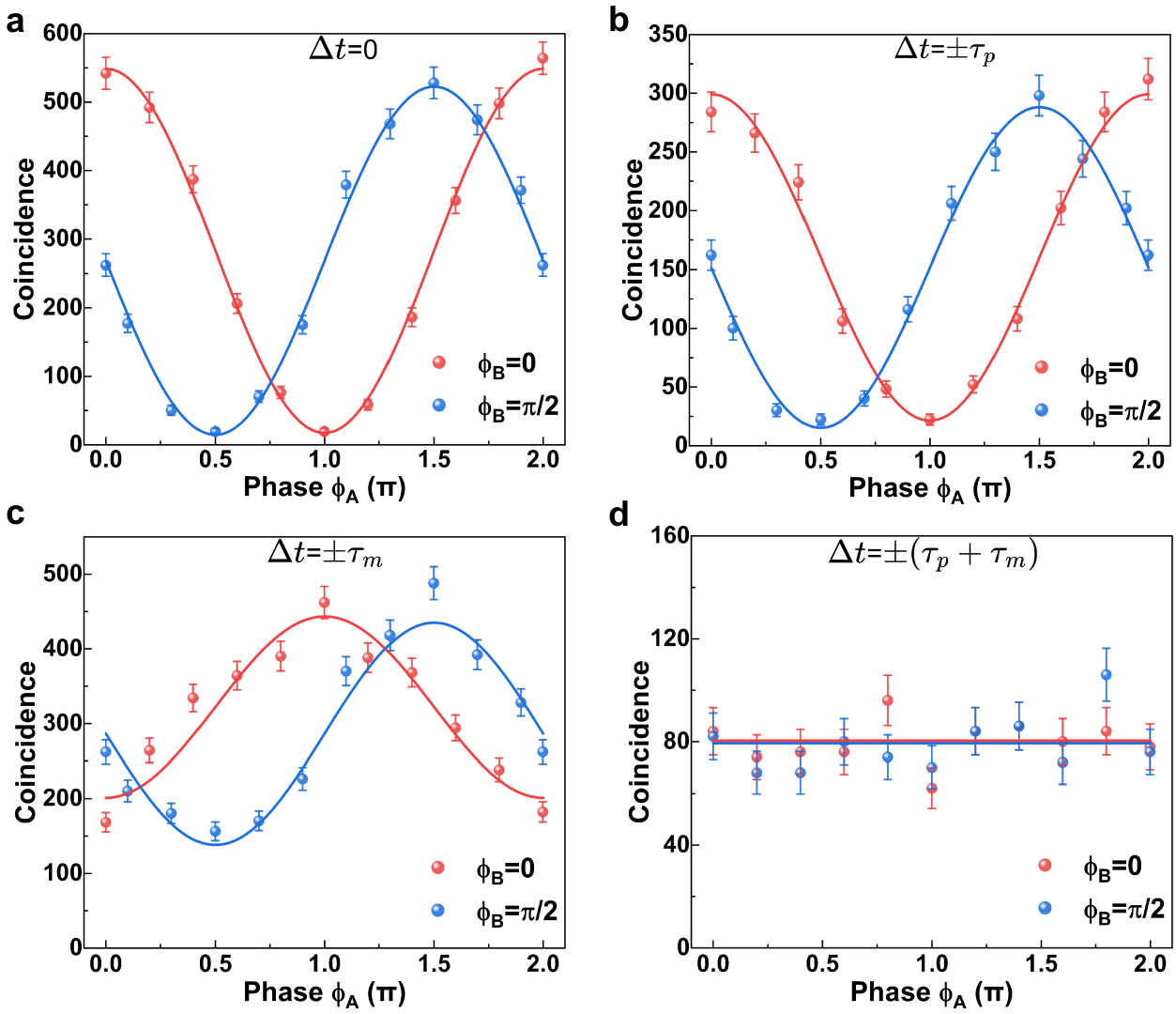}
\caption{\textbf{Franson interference fringes.} \textbf{a--d}, Interference curves of the coincidence counts $N(\phi_A,\phi_B)$ extracted at time delays of 0, $\pm\tau_p$, and $\pm\tau_m$, $\pm(\tau_p+\tau_m)$ respectively, with $\phi_B$ fixed at 0 and $\pi/2$. The solid dots represent the experimental data, and the solid lines are the fitting results.}
\label{fig4}
\end{figure}

We quantify the phase-dependent correlations by integrating the coincidence counts $N(\phi_A,\phi_B)$ over each characteristic delay peak, thereby extracting the one-dimensional fringes shown in Figs.~\ref{fig4}\textbf{a}--\ref{fig4}\textbf{d}. For each side characteristic delay, the plotted value is the sum 
of the two symmetric peaks at $\pm\tau_p$, $\pm\tau_m$, or $\pm(\tau_p+\tau_m)$. The data are fitted with sinusoidal functions consistent with Eqs.~(\ref{Center})--(\ref{3}).  Using two fit parameters $N_1$ (amplitude) and $N_2$ (offset),
the fringes at $\Delta t=0$ and $\pm\tau_p$ follow the sum phase,
$N_{0,\tau_p}=N_1[1+\cos(\phi_A+\phi_B)]+N_2$,
whereas the fringes at $\Delta t=\pm\tau_m$ follow the phase difference,
$N_{\tau_m}=N_1[5-4\cos(\phi_A-\phi_B)]+N_2$;
the outer peaks at $\pm(\tau_m+\tau_p)$ are fitted by a constant.
We obtain high-visibility interference at $\Delta t=0$ and $\pm\tau_p$, with $V_0=94.6\pm1.4\%$ and $V_{\tau_p}=86.6\pm1.6\%$, respectively, both exceeding the threshold $1/\sqrt{2}$ associated with Bell-type nonlocal correlations~\cite{tittel1998violation}. In contrast, the $\pm\tau_m$ peaks exhibit a reduced visibility of $V_{\tau_m}=51.8\pm3.1\%$, and the outer peaks show no measurable phase dependence.

\section*{Bell inequality violation}

To verify nonlocal quantum correlations, we evaluate the CHSH Bell parameter~\cite{Clauser1969PRL}, defined as
\begin{equation}
S=\left| E(\phi_A,\phi'_B)-E(\phi_A,\phi_B)+E(\phi'_A,\phi'_B)+E(\phi'_A,\phi_B)\right|,
\end{equation}
where the correlation function $E(\phi_A,\phi_B)$ is obtained from the coincidence counts between the two output ports at Alice and Bob,
\begin{equation}
E(\phi_A,\phi_B)=
\frac{N(\phi_A,\phi_B)+N(\phi_A^\perp,\phi_B^\perp)-N(\phi_A^\perp,\phi_B)-N(\phi_A,\phi_B^\perp)}{N(\phi_A,\phi_B)+N(\phi_A^\perp,\phi_B^\perp)+N(\phi_A^\perp,\phi_B)+N(\phi_A,\phi_B^\perp)}.
\end{equation}
Here, $\phi'_A$ ($\phi'_B$) and $\phi_A^\perp$ ($\phi_B^\perp$) denote the diagonal (shifted by $\pi/2$) and orthogonal (shifted by $\pi$) phase settings relative to $\phi_A$ ($\phi_B$), respectively. The quantity $N(\phi_A,\phi_B)$ refers to the coincidence counts for a given phase pair $(\phi_A,\phi_B)$, which can be extracted at different characteristic coincidence delays.
We set $(\phi_A,\phi_B,\phi'_A,\phi'_B)=(0,\pi/4,\pi/2,3\pi/4)$ and measure the corresponding 16 correlations. Because the two output ports at each node are complementary (differing by a phase of $\pi$), the required correlations can be obtained efficiently by integrating coincidences over the four detector-pair combinations $(A_1,B_1)$, $(A_1,B_2)$, $(A_2,B_1)$, and $(A_2,B_2)$ in a single acquisition.
By extracting the coincidence counts at $\Delta t=0$ and $\pm\tau_p$, we can compute the corresponding $S$ values for $\ket{2}$ and $\ket{11}$ contributions.

\begin{figure}
\centering 
\includegraphics[width=1\textwidth]{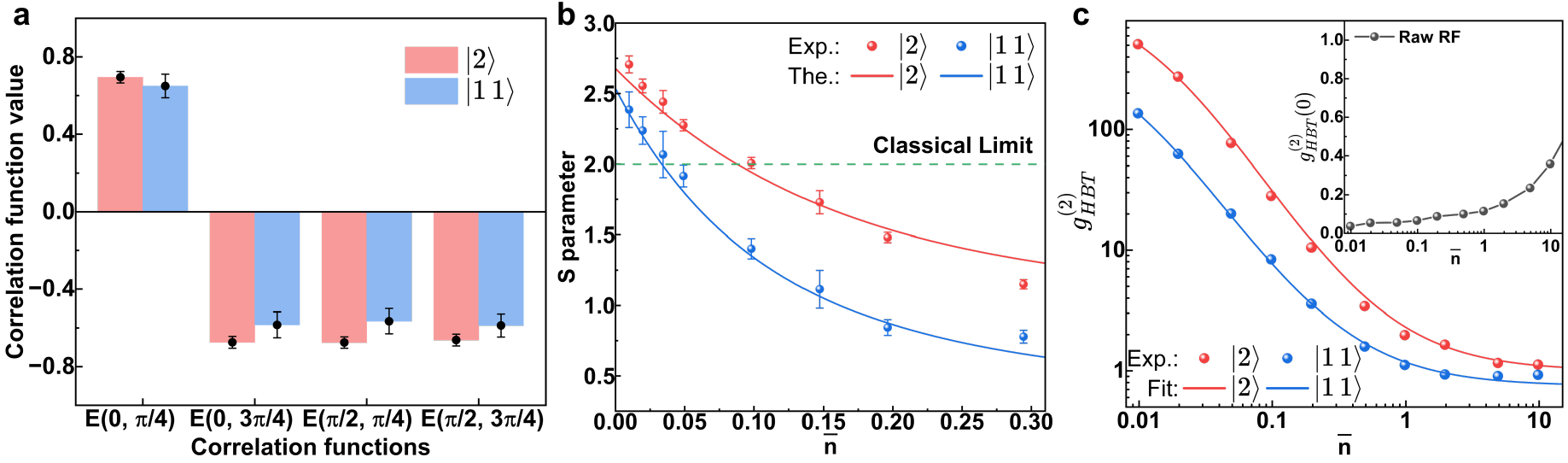}
\caption{\textbf{CHSH Bell inequality violation.}
\textbf{a}, Correlation functions at coincidence delays of $0$ and $\pm\tau_p$ for $\bar{n}=0.01$.
\textbf{b}, Excitation-power dependence of the CHSH $S$ parameters extracted at coincidence delays of $0$ (red) and $\pm\tau_p$ (blue).  \textbf{c}, Excitation-power dependence of the second-order autocorrelation $g^{(2)}_{\mathrm{HBT}}(0)$ of the AMZI-prepared output at $0$ (red) and $\pm\tau_p$ (blue) delays.  Inset: $g^{(2)}_{\mathrm{HBT}}(0)$ of  the raw RF versus $\bar{n}$.}
\label{fig5}
\end{figure} 

Figure~\ref{fig5}\textbf{a} shows the correlation functions measured at coincidence delays of $0$ and $\pm\tau_p$ for $\bar{n}=0.01$, yielding CHSH parameters $S_{0}=2.70\pm0.06$ and $S_{\pm\tau_p}=2.39\pm0.13$. Both values violate the classical bound ($S =2$), by 12 and 3 standard deviations respectively. Figure~\ref{fig5}\textbf{b} summarizes the excitation-power dependence: $S_0$ and $S_{\pm\tau_p}$ decrease with increasing $\bar{n}$. This degradation tracks the concomitant reduction of photon bunching in the AMZI-prepared field
(Fig.~\ref{fig5}\textbf{c}): when $S_0$ and $S_{\pm\tau_p}$ approach the classical limit at $\bar{n}=0.08$ and $0.04$, respectively, the bunching values at both delays fall to $\sim 40$. Notably, the single-photon purity of the raw RF  varies only slightly for $\bar{n}\le 0.3$, excluding significant degradation by residual laser background.

The observed power dependence follows directly from the RF state evolution with excitation strength: as $\bar{n}$ increases, the single-photon probability $p_1$ of the raw RF grows while $p_0$ decreases. In the ideal low-power limit, the HBT autocorrelations $g^{(2)}_{\mathrm{HBT}}(0)$ and $g^{(2)}_{\mathrm{HBT}}(\pm \tau_p)$ scale as $1/p_1^{2}$~\cite{wang2025coherence}. Using the experimental parameters, we develop a theoretical model that reproduces the measured power dependence of both the $S$ parameters and $g^{(2)}_{\mathrm{HBT}}(0)$, shown as solid lines in Figs.~\ref{fig5}\textbf{b} and \ref{fig5}\textbf{c}; details of the CHSH Bell-parameter evaluation are provided in Supplementary Information~1.5.

\section*{Conclusion}
In summary, we demonstrate simultaneously two types of two-photon entanglement from the RF of a QD via linear optics. By coherently reshaping the Heitler-regime signal with an AMZI, we suppress the single-photon component and synthesize two-photon states containing both simultaneous pairs $\ket{2}$ and temporally separated pairs $\ket{11}$. Franson measurements reveal high-visibility two-photon interference and clear CHSH violations, establishing RF as a reliable resource for generating nonlocal correlations using linear optics. Our source operates at ultralow excitation power and achieves a normalized brightness of $2.1\times10^9$~Hz/mW, as detailed in Supplementary Information~4, while remaining compatible with integrated photonic platforms.

The coexistence of $\ket{2}$ and $\ket{11}$ turns coincidence timing into a passive selector of two-photon projections, with $\Delta t=0$ and $\Delta t=\pm\tau_p$ accessing complementary components. Such time-resolved, passive basis selection could be leveraged in security-oriented settings to complicate intercept--resend attacks. Looking ahead, deterministic preparation of simultaneous photon pairs in the $\ket{2}$ sector, without admixture from time-separated components such as $\ket{11}$, may be achievable either by combining indistinguishable RF from two independent quantum emitters at a beam splitter, or by interfering the RF with an attenuated coherent field.

%%%%%%%%%%%%%%%% MATERIALS AND METHODS %%%%%%%%%%%%%%%

\section*{Methods}

\subsection*{QD-micropillar device}

The device was fabricated from a wafer grown by molecular beam epitaxy (MBE) on a semi-insulating GaAs substrate. An InAs quantum dot (QD) layer is embedded in the cavity region between upper and lower distributed Bragg reflectors (DBRs) comprising 18 and 30.5 pairs, respectively. The structure was processed into a $2.4~\mu\text{m}$-diameter micropillar. The sample was cooled to $6~\text{K}$ in a closed-cycle cryostat and tuned into resonance with the cavity mode at $\lambda = 911.470~\text{nm}$. We study the neutral-exciton transition, which has a linewidth of $2\pi\times 2.37~\text{GHz}$, corresponding to a radiative lifetime $T_1=67.2~\text{ps}$. The fine-structure splitting is $\sim 0.91~\text{GHz}$. The device operates in the weak-coupling regime with cooperativity $C=g^2/(\kappa\gamma)=F_p/2\simeq 5.45$, corresponding to a Purcell factor $F_p\simeq 10.9$. The cavity damping rate is $\kappa\simeq 35~\text{GHz}$, giving a quality factor $Q\simeq 9350$.

%%%%%%%%%%%%%%%% MAIN TEXT TABLES %%%%%%%%%%%%%%%

%%%%%%%%%%%%%%%% REFERENCES %%%%%%%%%%%%%%%

\bibliography{mainref} 
\bibliographystyle{\natphysbibstyle}

%%%%%%%%%%%%%%%% ACKNOWLEDGEMENTS %%%%%%%%%%%%%%%

\section*{Acknowledgments}

The authors acknowledge C. Sanchez Muñoz and J. Feist for help on a formal simulation approach for the circuit in Fig.~2. This work was supported by the National Natural Science Foundation of China under grants (12494604, 12494600, 12204049, and 12504410), the Innovation Program for Quantum Science and Technology (2024ZD0302500 and 2021ZD0300801), and
the Beijing Postdoctoral Research Foundation. C.A-S. acknowledges the support from the projects from the Ministerio de Ciencia e Innovaci\'on PID2023-148061NB-I00 and PCI2024-153425, the project ULTRABRIGHT from the Fundaci\'on Ram\'on Areces, the Grant “Leonardo for researchers in Physics 2023” from Fundaci\'on BBVA, and the support from the Comunidad de Madrid fund “Atracci\'on de Talento, Mod. 1”, Ref. 2020-T1/IND-19785.

%%%%%%%%%%%%%%%% SUPPLEMENT LIST %%%%%%%%%%%%%%%

% List the contents of your Supplementary Materials, including the numbers of any
% supplementary figures, tables, external data files etc. and any references that are
% cited only in the supplement. In this example, refs. 7-8 are cited only in the supplement.
% Fill out your numbers accordingly and delete any lines that aren't applicable.
%\subsection*{Supplementary Information\\
%Supplementary Information\\
%Figs. S1 to S7\\
%Tables S1\\
%References~\cite{dalibard1983correlation,hohn2023energy,zhao2020high,finco2024time,lu2019chip} are cited only in the supplement.
%References (38-42) are cited only in the supplement.
%References (37-38)% automatically fills out the last reference number
% (filling out the other numbers automatically is possible but fiddly and liable to break)
\clearpage
\fi
\ifshowSI
    \newpage

%%%%%%%%%%%%%%%% END OF MAIN TEXT %%%%%%%%%%%%%%%

%%%%%%%%%%%%%%%% START OF SUPPLEMENT %%%%%%%%%%%%%%%

% Figures, tables, equations and pages in the supplement are numbered S1, S2 etc.
\renewcommand{\thefigure}{S\arabic{figure}}
\renewcommand{\thetable}{S\arabic{table}}
\renewcommand{\theequation}{S\arabic{equation}}
\renewcommand{\thepage}{S\arabic{page}}
\setcounter{figure}{0}
\setcounter{table}{0}
\setcounter{equation}{0}
\setcounter{page}{1} % not 0 as \newpage already started a supplementary page
% References continue the numbering from the main text.

%%%%%%%%%%%%%%%% SUPPLEMENT TITLE PAGE %%%%%%%%%%%%%%%

\begin{center}
\section*{Supplementary Information for\\ \scititle}

% Author list for the supplement
% Indicate the corresponding authors, but do NOT include institutions here
% It would be nice if the template auto-generated this, but doing so is complicated...
	Jian~Wang$^{1\dagger}$,
	Xiu-Bin Liu$^{1,2\dagger}$,
    Ziqi Zeng$^{1,3,4}$,
	Xu-Jie Wang$^{1}$,
    Carlos Antón-Solanas$^{5,6}$ \\
    Li Liu$^{1}$,
    Hanqing~Liu$^{7,8}$,   
    Haiqiao~Ni$^{7,8}$, 
    Zhichuan~Niu$^{7,8}$,
    Bang Wu$^{1}$,
    Zhiliang~Yuan$^{1\ast}$
\\ % we're not in a \author{} environment this time, so use \\ for a new line
\small$^\ast$Corresponding author. Email: yuanzl@baqis.ac.cn\\
\small$^\dagger$These authors contributed equally to this work.
\end{center}

%%%%%%%%%%%%%%%% Supplementary Information %%%%%%%%%%%%%%%
\newpage

\renewcommand{\thesection}{\arabic{section}}
\renewcommand{\thesubsection}{\arabic{section}.\arabic{subsection}}
\renewcommand{\thesubsubsection}{\arabic{section}.\arabic{subsection}.\arabic{subsubsection}}

\tableofcontents

\section{Theory}

\subsection {Introduction}

\begin{figure}[htb]
\centering 
\includegraphics[width=0.85\textwidth]{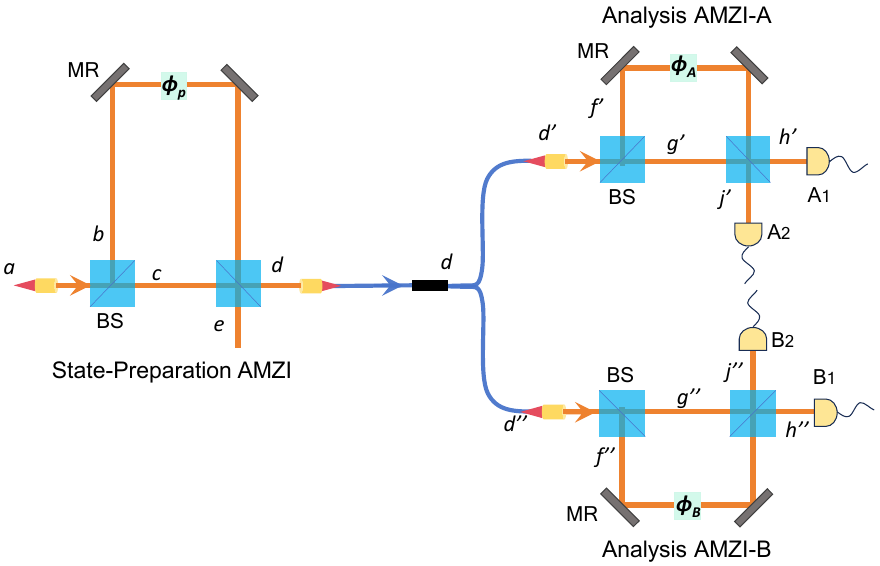}
\caption{\textbf{Simplified schematic of the interferometric setup.} The state-preparation AMZI (left, operated at $\phi_p=\pi$) receives the resonance-fluorescence field at port $a$ and prepares, at port $d$, the two-photon states used in the experiment. The output is then split and analyzed by two analysis AMZIs (right) with independently controllable phases $\phi_A$ and $\phi_B$. All relevant optical ports are labeled. AMZI: asymmetric Mach-Zehnder interferometer; BS: beam splitter; MR: mirror.}
\label{S1}
\end{figure}

Figure~\ref{S1} shows the interferometric network used in the theoretical analysis; the italic letters
(\textit{a}--\textit{j}$''$) label spatial ports or modes.
The RF field is injected at port \textit{a} and enters the state-preparation asymmetric Mach-Zehnder interferometer (AMZI).
Its short and long arms (paths \textit{c} and \textit{b}, respectively) are unbalanced by a delay $\tau_p$, and a phase shifter in the long arm sets the relative phase $\phi_p$.
The prepared field exits at port \textit{d} and is sent to a 50:50 fiber beam splitter (FBS), which produces two equal outputs, \textit{d}$'$ and \textit{d}$''$, that feed two identical analysis AMZIs forming a Franson-type configuration~\cite{franson1989bell}.
The arm-delay imbalance $\tau_m$ of the analysis AMZIs is intentionally chosen to be different from the state-preparation imbalance $\tau_p$, so that interference associated with the state-preparation AMZI is temporally separated and does not contribute to (or obscure) the Franson analysis.
The phases $\phi_A$ and $\phi_B$ in the upper (Alice) and lower (Bob) analysis AMZIs are independently adjustable.
After recombination at the second beam splitter of each analyzer, the two output ports \textit{h}$'$--\textit{j}$'$ (Alice) and \textit{h}$''$--\textit{j}$''$ (Bob) are detected by $A_1/A_2$ and $B_1/B_2$, respectively, for coincidence measurements.

In the remainder of this theory subsection, we follow the port (mode) labels in Fig.~\ref{S1} throughout. Unless otherwise stated, each field operator carries a subscript indicating its temporal mode (detection time) and a spatial label corresponding to the interferometer port shown in Fig.~\ref{S1}. All mode transformations are therefore written explicitly as input--output relations between these labeled ports, and the same labeling convention is used consistently in the derivations of the coincidence amplitudes and probabilities.

\subsection{The resonance-fluorescence input state}
\label{sec:RF_state}

Before analyzing the state-preparation AMZI, we specify the quantum state of the resonance-fluorescence (RF) emission. Following the pure-state framework~\cite{wang2025coherence}, we treat the coherently driven two-level emitter and its radiation as a joint quantum system. Within a temporal mode $t$, the joint state is written as
\begin{align}
\ket{\Psi_t}
=
\sqrt{p_0}\,\ket{g,0}
+
\sqrt{p_1}\,e^{-i\omega_L t}\frac{\ket{e,0}+\ket{g,1}}{\sqrt{2}},
\end{align}
where $\ket{g}$ and $\ket{e}$ denote the ground and excited states of the emitter, $\ket{0}$ and $\ket{1}$ are the vacuum and single-photon states of the emitted field, and $\omega_L$ is the laser angular frequency.

Tracing over the emitter ($\ket{g}$ and $\ket{e}$) yields a mixed photonic state. Nevertheless, this state is equivalent to the photon-number coherent state
\begin{align}
\ket{\psi}
=
\sqrt{p_0}\ket{0}
+
\sqrt{p_1}\,e^{-i\omega_L t}\ket{1},
\label{eq:photon_state}
\end{align}
followed by an effective $50\%$ channel attenuation. As shown in Ref.~\cite{wang2025coherence}, such loss does not modify the relevant interference observables. We therefore use the pure photonic state in Eq.~\eqref{eq:photon_state} for the derivations below.

For a sequence of temporal modes separated by delays much longer than the emitter's coherence time, the RF input state is well approximated by a tensor product,
\begin{align}
\ket{\psi_{\mathrm{in}}}
=
\bigotimes_t \ket{\psi_t}, \label{eq:inputstate}
\end{align}
which forms the input to the state-preparation AMZI.

\subsection{The quantum state prepared by the state-preparation AMZI}

We derive the quantum state prepared by the state-preparation AMZI. 
Owing to the delay imbalance, the output must be described jointly in two consecutive temporal modes separated by $\tau_p$, which we take to be centered at times $t$ and $t-\tau_p$. The output field in the temporal mode at $t$ receives contributions from the input at port $a$ at time $t$ (short arm) and at $t-\tau_p$ (long arm). Likewise, the output temporal mode at $t-\tau_p$ is formed from the input at $t-\tau_p$ and at $t-2\tau_p$. Consequently, describing the two output temporal modes at $t$ and $t-\tau_p$ requires only the input state over three successive emission times: $t$, $t-\tau_p$, and $t-2\tau_p$. 
According to Eq.~\ref{eq:inputstate}, the input state at port $a$ over three consecutive temporal modes can be written as a tensor product,
\begin{align}
\ket{\psi_{in}} & =|\psi\rangle_{t-2\tau_p}\,|\psi\rangle_{t-\tau_p}\,|\psi\rangle_{t} \notag \\
& =
p_0^{3/2}\ket{000} + p_0\sqrt{p_1}\left(\ket{001} + \ket{010} + \ket{100}\right)+ \sqrt{p_0}p_1\left(\ket{011} + \ket{101} + \ket{110}\right)+ p_1^{3/2}\ket{111}.
\label{Eq.in2}
\end{align}
Here, the global excitation phase of the driving laser is omitted for simplicity. 

Below, we first adopt a simplified treatment to highlight the physical action of the state-preparation AMZI, and then present a rigorous calculation yielding the exact output state at port $d$.

\subsubsection{Simplified analysis}

Since our interest is in the output state spanning two adjacent temporal modes, $t-\tau_p$ and $t$, we begin by writing the input--output relations for the field annihilation operators of the state-preparation AMZI. In what follows, we retain only the operators associated with these two temporal modes (and the preceding input mode at $t-2\tau_p$ that feeds into $t-\tau_p$), and for simplicity we neglect losses at the first beam splitter. Under these assumptions, the mode transformation reads
\begin{align}
\begin{split}
a_{t-2\tau_p} &= \frac{1}{\sqrt{2}} \big[ d_{t-\tau_p} + e_{t-\tau_p} \big] \,,\\
a_{t-\tau_p} &= \frac{1}{\sqrt{2}} \Big\{ [d_t + e_t] + e^{-i\phi_p} [d_{t-\tau_p} - e_{t-\tau_p}] \Big\} \,,\\
a_{t} &= \frac{1}{\sqrt{2}} e^{-i\phi_p} [d_t - e_t] \,.
\end{split}
\end{align}

By applying these transformations to the input state $\ket{\psi_{in}}$ in Eq.~\ref{Eq.in2}, one obtains the normalized joint output quantum state at ports $d$ and $e$ for the temporal modes $t-\tau$ and $t$, which is given by
%{\small
\begin{align}
\begin{aligned}
\ket{\psi_{\text{out}}} = &\ p_0^{3/2} \ket{00} \\
&+ \frac{e^{-i\phi_p} p_0 \sqrt{p_1}}{\sqrt{2}} \left[ (1 + e^{i\phi_p}) \ket{01_d} +(-1 + e^{i\phi_p}) \ket{01_e}+(1 + e^{i\phi_p}) \ket{1_d 0}+(-1 + e^{i\phi_p}) \ket{1_e 0}\right]  \\
&+ \frac{e^{-i\phi_p} \sqrt{p_0} p_1}{\sqrt{2}} \left( \ket{02_d } - \ket{02_e} + \ket{2_d 0} - \ket{2_e 0} \right) \\
&+ \frac{e^{-2 i\phi_p} \sqrt{p_0} p_1}{2} [ (1 + e^{i\phi_p} + e^{2 i\phi_p}) \ket{1_d 1_d} + (1 - e^{i\phi_p} + e^{2 i\phi_p}) \ket{1_e 1_e} \\
&+(-1 - e^{i\phi_p} + e^{2 i\phi_p}) \ket{1_d 1_e} +(-1 + e^{i\phi_p} + e^{2 i\phi_p}) \ket{1_e 1_d}  ]\\
&+ \frac{ e^{-i\phi_p} p_1^{3/2}}{2} \left( \ket{1_d 2_d} - \ket{1_d 2_e}+\ket{1_e 2_d} - \ket{1_e 2_e} \right)+ \frac{e^{-2 i\phi_p} p_1^{3/2} }{2} \left( -\ket{2_d 1_e} + \ket{2_d 1_d}+\ket{2_e 1_e} - \ket{2_e 1_d} \right),
\label{out.1}
\end{aligned}
\end{align}
%}

Here, each two-mode Fock ket refers to the two temporal bins $(t-\tau)$ and $t$. 
Specifically, a basis state written as $\ket{n_{x}\, m_{y}}$ (or $\ket{n_{x}0}$, $\ket{0\, m_{y}}$) means:
the first entry gives the photon number in the early bin $(t-\tau)$ exiting from port $x\in\{d,e\}$,
and the second entry gives the photon number in the late bin $t$ exiting from port $y\in\{d,e\}$.
For example, $\ket{1_d 0}$ denotes one photon in the $(t-\tau)$ bin at port $d$ and vacuum in the $t$ bin, 
while $\ket{0\,1_e}$ denotes vacuum in the $(t-\tau)$ bin and one photon in the $t$ bin at port $e$.
A ``0'' in either entry denotes vacuum in that temporal bin, i.e., no photon exits from either output port in that bin.

We use Eq.~\eqref{out.1} to elucidate how the state-preparation AMZI acts as a phase-tunable filter on the RF field. 
Reading the state line by line: (i) the first line is the vacuum sector, (ii) the second line is the single-photon sector, 
(iii) the third line describes the two-photon Fock states $\ket{2}$ ($\ket{20}$ or $\ket{02}$), (iv) the fourth and fifth lines is the time-bin two-photon states $\ket{1_{d,e}1_{d,e}}$, and (v) the final line contains the three-photon sector. 
When the AMZI is phase-randomised,  the  probability weight in each photon-number sector scales as 
$P_{0}\sim p_0^{3}$, $P_{1}\sim p_0^{2}p_1$, $P_{2}\sim p_0 p_1^{2}$, and $P_{3}\sim p_1^{3}$. However, once $\phi_p$ can be set deterministically, the single-photon amplitude in a chosen output port (either $d$ or $e$) becomes fully controllable: it can be tuned continuously from its maximum value to complete extinction via destructive interference. 

In our experiment, we set $\phi_p=\pi$ so that the direct single-photon amplitudes at port $d$---namely the $\ket{01_d}$ and $\ket{1_d0}$ terms in the second line of Eq.~\eqref{out.1}---vanish due to destructive interference.
Notably, this does not mean that port $d$ is completely free of single-photon contributions: they re-enter at higher order via the two-photon sector (fourth line).
For instance, the $\ket{1_d1_e}$ and $\ket{1_e1_d}$ terms correspond to one photon in port $d$ together with a second photon in port $e$; when one focuses on port $d$ alone (i.e., ignores port $e$), these states contribute an effective single-photon component in port $d$. 
The key point is the scaling: the suppressed single-photon sector in the second line carries probability that scales with  $\mathcal{O}(p_1)$, whereas the residual single-photon component inherited from the fourth line carries probability scaling $\mathcal{O}(p_1^2)$. 
Therefore, in the weak-excitation limit $p_1\to 0$, the output in port $d$ is dominated by vacuum, while the desired two-photon terms provide the leading contribution to two-photon coincidence measurements—the residual single-photon background contributes only to accidental coincidences at a much lower order ($\mathcal{O}(p_1^4)$).

\subsubsection{Complete calculation.}

We now present a rigorous and complete calculation of the state-preparation AMZI transformation. The relations between the input and output field operators of the AMZI can be written as
\begin{align} 
\begin{split} 
a_{t-2\tau_p} &= \frac{1}{2} \big\{ [d_{t-\tau_p} + e_{t-\tau_p}]
+ e^{-i\phi_p} [d_{t-2\tau_p} - e_{t-2\tau_p}]\big\} \,,\\ 
a_{t-\tau_p} &= \frac{1}{2} \big\{ [d_t + e_t]
+ e^{-i\phi_p} [d_{t-\tau_p} - e_{t-\tau_p}] \big\} \,,\\ 
a_{t} &= \frac{1}{2}\big\{ [d_{t+\tau_p} + e_{t+\tau_p}]
+ e^{-i\phi_p} [d_t - e_t] \big\} \,.
\label{tansf.2}
\end{split} 
\end{align}

By applying the above operator transformations to the input state given in Eq.~\ref{Eq.in2}, we obtain the normalized joint output quantum state at ports $d$ and $e$. Each output port involves four temporal modes, namely $t-2\tau$, $t-\tau$, $t$, and $t+\tau$. Consequently, the full output state must be represented in an eight-mode Fock basis. For example, the state $\ket{00000000}$ denotes vacuum in all four temporal modes at port $d$ (first four entries) and port $e$ (last four entries), respectively. The complete analytical expression of the output state is lengthy, comprising a total of 85 distinct quantum-state components. For brevity and clarity, we only show partial results,
\begin{align}
\ket{\psi_{\text{out}}}
&= p_0^{3/2}\ket{00000000}
- \frac{p_0\sqrt{p_1}}{2}\ket{00000001}
+ \frac{(-1+e^{i\phi_p})p_0\sqrt{p_1}}{2}\ket{00000010}
\notag\\
&\quad
+ \frac{\sqrt{p_0}p_1}{4}\ket{00000011}
- \frac{e^{i\phi_p}\sqrt{p_0}p_1}{2\sqrt{2}}\ket{00000020}
+ \frac{(-1+e^{i\phi_p})p_0\sqrt{p_1}}{2}\ket{00000100}
\notag\\
&\quad
+ \frac{e^{i\phi_p}(-1+2\cos\phi_p)\sqrt{p_0}p_1}{4}\ket{00000110}
- \frac{p_1^{3/2}}{8}\ket{00000111}
+ \frac{e^{i\phi_p}p_1^{3/2}}{4\sqrt{2}}\ket{00000120}
\notag\\
&\quad
- \frac{e^{i\phi_p}\sqrt{p_0}p_1}{2\sqrt{2}}\ket{00000200}
+ \cdots.
\label{out.2}
\end{align}

By sequentially tracing out the two incomplete temporal modes $t-2\tau$ and $t+\tau$, as well as all modes associated with output port $e$, from the output state given in Eq.~\ref{out.2}, we obtain the reduced density matrix describing the field at port $d$ in the two relevant temporal modes $t-\tau$ and $t$. The reduced density matrix is expressed in the Fock basis $ {|0,0\rangle, |0,1\rangle, |0,2\rangle, |1,0\rangle, |1,1\rangle, |1,2\rangle, |2,0\rangle, |2,1\rangle }$, where the two indices denote the photon numbers in the $t-\tau_p$ and $t$ modes, respectively.The diagonal elements of the density matrix,
$\rho_{n_1 n_2} = \langle n_1, n_2 | \rho | n_1, n_2 \rangle$, are given by the following expressions:
\begin{align}
\begin{aligned}
\rho_{00} &= \frac{1}{16} \left[ (p_0+3)(p_0^2+p_0+2) - 4p_0 p_1 (p_0+3) \cos\phi_p + 2 p_0 p_1^2 \cos(2\phi_p) \right] , \\
\rho_{01} &= \rho_{10} = \frac{p_1}{32} \left[ (p_0+1)(3p_0+5) + 8p_0(p_0+1) \cos\phi_p - 4p_0 p_1 \cos(2\phi_p) \right] , \\
\rho_{02} &= \rho_{20} = \frac{1}{32} p_1^2 (p_0+3) , \\
\rho_{11} &= \frac{p_1^2}{16} \left[ (2p_0+1) + 4p_0 \cos\phi_p + 2p_0 \cos(2\phi_p) \right] , \\
\rho_{12} &= \rho_{21} = \frac{p_1^3}{32}.
\end{aligned}
\end{align}

%Substituting $\phi_p = \pi$ into the above expressions and further simplifying, 
Using the experimental setting $\phi_p = \pi$, we obtain
\begin{align}
\begin{aligned}
\rho_{00} &= 1 + \frac{(-11 + p_1)p_1^2}{16} \overset{p_1 \ll p_0 }{\rightarrow}  1-\frac{11 p_1^2}{16} , \\
\rho_{01}  &= \rho_{10} = -\frac{(-6 + p_1)p_1^2}{32} \overset{p_1 \ll p_0 }{\rightarrow} \frac{3 p_1^2}{16}, \\
\rho_{02}  &= \rho_{20} = -\frac{(-4 + p_1)p_1^2}{32} \overset{p_1 \ll p_0 }{\rightarrow}\frac{p_1^2}{8}, \\
\rho_{11} &= \frac{p_1^2}{16} , \\
\rho_{12}  &= \rho_{21} = \frac{p_1^3}{32}.
\end{aligned}
\end{align}

Under weak excitation ($p_1 \to 0$), the vacuum population $\rho_{00}$ overwhelmingly dominates the output state at port $d$.
Among the non-vacuum terms, the populations of the single-photon components ($\rho_{01}$, $\rho_{10}$) and the two-photon components ($\rho_{02}$, $\rho_{20}$, and $\rho_{11}$) are of the same order, scaling as $p_1^{2}$, and they are all much larger than the three-photon contributions ($\rho_{12}$, $\rho_{21}$), which scale as $p_1^{3}$.

\subsection{Coincidence probabilities at the characteristic delays}

\subsubsection{Franson coincidence-delay histograms}

When the AMZI-prepared state is injected into the Franson interferometer (Fig.~\ref{S1}), a coincidence requires two photons within the coincidence window. 
The recorded Alice--Bob coincidence histogram is therefore naturally organized in the coincidence-delay variable $\Delta t$, which encodes the relative path choices taken in all the AMZIs. 
Figure~\ref{S2}(a) summarizes the coincidence-delay structure expected in a Franson measurement when the imbalance of the analysis AMZIs, $\tau_m$, differs from that of the state-preparation AMZI, $\tau_p$. 
In the general case, the Alice--Bob coincidence histogram contains nine well-separated peaks at $0$, $\pm\tau_m$, $\pm\tau_p$, $\pm(\tau_p-\tau_m)$, and $\pm(\tau_p+\tau_m)$.
Also evident from Fig.~\ref{S2}(a) is that phase-dependent interference occurs only in the central peaks at $\Delta t=0$ and $\Delta t=\pm\tau_p$. This is because only these peaks collect events for which two alternative two-photon pathways---namely, both photons taking the short arms or both taking the long arms of the analysis AMZIs---are temporally indistinguishable and therefore interfere. 
The $\Delta t=0$ peak is contributed by the $\ket{2}$ sector (either $\rho_{20}$ or $\rho_{02}$): both photons are registered in the same time bin, and the two indistinguishable analysis alternatives $(A_s,B_s)$ and $(A_\ell,B_\ell)$ yield the same coincidence-time difference and therefore interfere with a phase set by $\phi_A$ and $\phi_B$. 
The peaks at $\Delta t=\pm\tau_p$ are contributed by the $\ket{11}$ sector (one photon in each time bin), where the delay $\pm\tau_p$ reflects the intrinsic time-bin separation between the two photons; within each peak, the same two indistinguishable analysis alternatives $(A_s,B_s)$ and $(A_\ell,B_\ell)$ again lead to the same measured delay and therefore interfere. 
In contrast, all satellite peaks surrounding $0$ and $\pm\tau_p$ (i.e., those at $\pm\tau_m$, $\pm(\tau_p-\tau_m)$, and $\pm(\tau_p+\tau_m)$) originate from mismatched-arm contributions $(A_s,B_\ell)$ or $(A_\ell,B_s)$, for which the coincidence-time differences are distinguishable; the corresponding amplitudes do not overlap in the coincidence-delay domain, and hence no interference visibility is expected.

Figure~\ref{S2}(b) shows the special case realized in our experiment, $\tau_p=2\tau_m$. 
In the general situation $\tau_p\neq 2\tau_m$, the contributions associated with the delays $\pm\tau_m$ and $\pm(\tau_p-\tau_m)$ appear at different coincidence delays and are therefore temporally distinguishable; their amplitudes do not overlap in the $\Delta t$ domain and no interference can occur between them. 
By contrast, when $\tau_p=2\tau_m$ one has $\tau_p-\tau_m=\tau_m$, so the nominal satellite peaks at $\pm\tau_m$ and $\pm(\tau_p-\tau_m)$ become degenerate and merge, reducing the nine-peak pattern to seven distinct peaks. 
This degeneracy removes the temporal distinguishability between the two contributions at $\Delta t=\pm\tau_m$, allowing their amplitudes to overlap and interfere, which produces a phase-dependent modulation at $\pm\tau_m$ in this special configuration. 
Because the two overlapping contributions generally have unequal amplitudes, their interference is incomplete and the resulting visibility at $\pm\tau_m$ is finite rather than unity, consistent with the modified contrasts observed in Fig.~\ref{S2}(b).

\begin{figure}[htb]
\centering 
\includegraphics[width=0.85\textwidth]{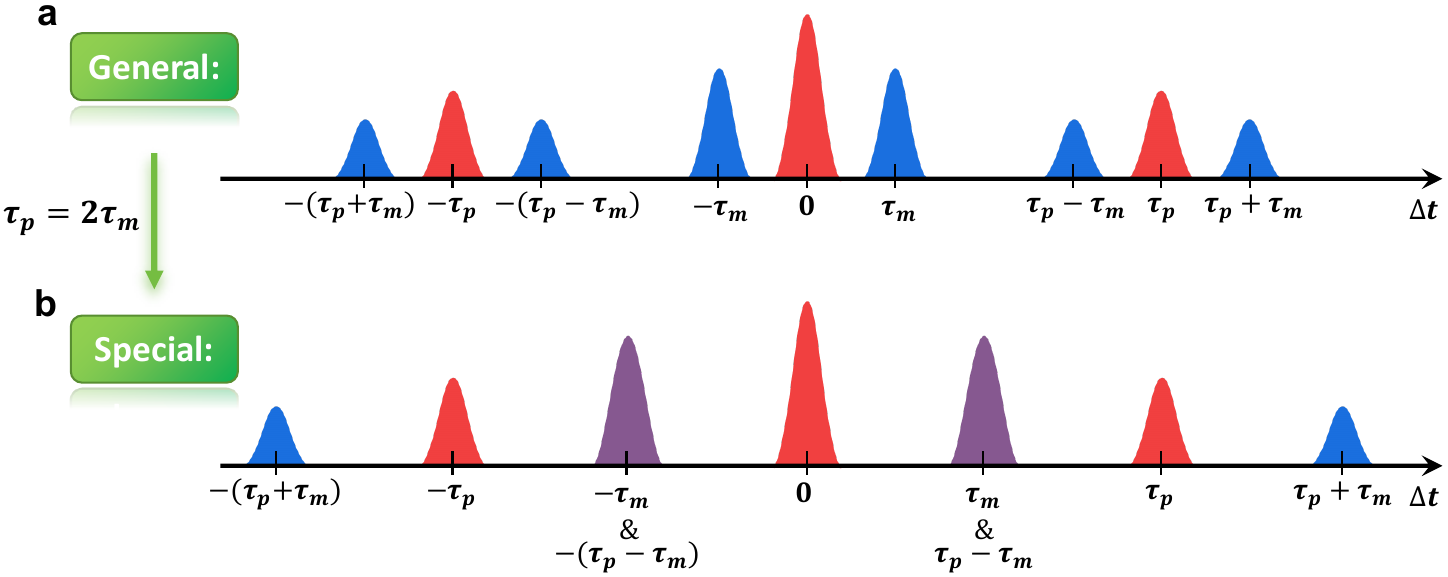}
\caption{\textbf{Schematic Franson coincidence-delay histograms.} \textbf{a}, General configuration; \textbf{b}, Experimental configuration (special case $\tau_p = 2\tau_m$). Peak amplitudes are illustrative only and do not represent actual values. The illustrations are color coded: red indicates phase-dependent peak amplitudes (high visibility), blue indicates phase-independent amplitudes, and purple indicates phase-dependent amplitudes with reduced interference visibility. 
}
\label{S2}
\end{figure}

\subsubsection{Coincidence operator $h_t^\prime h_{t+\Delta t}^{\prime\prime}$}

For brevity, we derive the coincidence probability for the Alice--Bob \textit{h}$^\prime$--\textit{h}$^{\prime\prime}$ port pair; the results for the other three output-port combinations follow analogously.
We define a coincidence event by the time tags of the two detection clicks: $t_1$ is the arrival time of a photon registered at output port $h'$, and $t_2$ is the arrival time of a photon registered at output port $h''$. 
The coincidence delay is then defined as $\Delta t = t_2 - t_1$, i.e., the time interval between the two clicks.
 For the sake of simplicity and completeness, we employ a reverse-calculation method in theoretical calculations. Using the unitary transformation relations for the annihilation operators across all AMZIs, we derive the explicit operator forms corresponding to the detection modes $h^{\prime}$ and $h^{\prime\prime}$.
\begin{align}
h_t^{\prime} &= \frac{1}{\sqrt{2}}\left(e^{i \phi_A} f_{t-\tau_m}^{\prime} + g_t^{\prime}\right) = \frac{1}{2}\left(e^{i \phi_A} d_{t-\tau_m}^{\prime} + d_t^{\prime}\right) \notag \\
&= \frac{1}{4}\left[e^{i \phi_A} \left(b_{t-\tau_p-\tau_m} + c_{t-\tau_m}\right) + \left(e^{i \phi_p} b_{t-\tau_p} + c_t\right)\right] \notag \\
&= \frac{1}{4\sqrt{2}}\left[e^{i \phi_A}\left(e^{i \phi_p} a_{t-\tau_p-\tau_m} + a_{t-\tau_m}\right) + \left(e^{i \phi_p} a_{t-\tau_p} + a_t\right)\right] \notag \\
&= \frac{1}{4\sqrt{2}}\left[e^{i(\phi_A+\phi_p)} a_{t-\tau_p-\tau_m} + e^{i \phi_A} a_{t-\tau_m} + e^{i \phi_p} a_{t-\tau_p} + a_t\right].
\\
h_{t+\Delta t}^{\prime\prime} &= \frac{1}{4\sqrt{2}} \left[ e^{i\phi_B} \left(e^{i\phi_p} a_{t+\Delta t-\tau_p-\tau_m} + a_{t+\Delta t-\tau_m} \right) + \left(e^{i\phi_p} a_{t+\Delta t-\tau_p} + a_{t+\Delta t} \right) \right] \notag \\
&= \frac{1}{4\sqrt{2}} \left[ e^{i(\phi_B+\phi_p)} a_{t+\Delta t-\tau_p-\tau_m} + e^{i\phi_B} a_{t+\Delta t-\tau_m} + e^{i\phi_p} a_{t+\Delta t-\tau_p} + a_{t+\Delta t} \right].
\end{align}
With the explicit operator forms for ports $h^{\prime}$ and $h^{\prime\prime}$ established, the coincidence probability between the two detectors is obtained from the normally ordered expectation value of the field operators,
\begin{equation}
\mathcal{C}=\left\langle\psi_{\mathrm{in}}\right|  \hat{h}^{\prime \prime \dagger}_{t+\Delta t} \hat{h}^{\prime \dagger}_t \hat{h}^{\prime}_t \hat{h}^{\prime \prime}_{t+\Delta t} \left|\psi_{\mathrm{in}}\right\rangle,
\end{equation}
where 
\begin{align}
h_t^{\prime} h_{t+\Delta t}^{\prime \prime} &= \frac{1}{32}\left[e^{i\left(\phi_A+\phi_B+2 \phi_p\right)} a_{t-\tau_p-\tau_m} a_{t+\Delta t-\tau_p-\tau_m}+e^{i\left(\phi_A+\phi_B+\phi_p\right)} a_{t-\tau_p-\tau_m} a_{t+\Delta t-\tau_m}\right. \notag \\
&\quad +e^{i\left(\phi_A+2 \phi_p\right)} a_{t-\tau_p-\tau_m} a_{t+\Delta t-\tau_p}+e^{i\left(\phi_A+\phi_p\right)} a_{t-\tau_p-\tau_m} a_{t+\Delta t} \notag \\
&\quad +e^{i\left(\phi_A+\phi_B+\phi_p\right)} a_{t-\tau_m} a_{t+\Delta t-\tau_p-\tau_m}+e^{i\left(\phi_A+\phi_{B}\right)} a_{t-\tau_m} a_{t+\Delta t-\tau_m} \notag \\
&\quad +e^{i\left(\phi_A+\phi_p\right)} a_{t-\tau_m} a_{t+\Delta t-\tau_p}+e^{i \phi_A} a_{t-\tau_m} a_{t+\Delta t} \notag \\
&\quad +e^{i\left(\phi_B+2 \phi_p\right)} a_{t-\tau_p} a_{t+\Delta t-\tau_p-\tau_m}+e^{i\left(\phi_B+\phi_p\right)} a_{t-\tau_p} a_{t+\Delta t-\tau_m} \notag \\
&\quad +e^{2 i \phi_p} a_{t-\tau_p} a_{t+\Delta t-\tau_p}+e^{i \phi_p} a_{t-\tau_p} a_{t+\Delta t} \notag \\
&\quad \left.+e^{i\left(\phi_B+\phi_p\right)} a_t a_{t+\Delta t-\tau_p-\tau_m}+e^{i \phi_B} a_t a_{t+\Delta t-\tau_m}+e^{i \phi_p} a_t a_{t+\Delta t-\tau_p}+a_t a_{t+\Delta t}\right].
\label{Eq.1}
\end{align}

Equation~\ref{Eq.1} implies that, for $\tau_p=2\tau_m$, a coincidence at detection-time interval $\Delta t \equiv t_2-t_1$ can arise from two clicks at times $t_1=t$ and $t_2=t+\Delta t$, whose associated photons could have been generated at port $a$ in temporal modes drawn respectively from
\[
t-3\tau_m,\; t-2\tau_m,\; t-\tau_m,\; t,
\]
and
\[
t+\Delta t-3\tau_m,\; t+\Delta t-2\tau_m,\; t+\Delta t-\tau_m,\; t+\Delta t.
\]

For a given interval $\Delta t$, the coincidence event can generally be produced by several distinct alternatives—different emission-time assignments at port $a$ together with different short/long path choices through the AMZIs—that nevertheless yield the same pair of detection times $(t,\;t+\Delta t)$. 
When these alternatives are indistinguishable by timing, they add coherently and produce two-photon interference.  As a result, interference occurs for certain characteristic delays (notably $\Delta t=0,\ \pm\tau_p$).
In the following, we analyze the coincidence contributions for the relevant $\Delta t$ values and identify the physical origin of the observed peaks.

\subsubsection{$\Delta t = 0$}

For $\Delta t=0$, the coincidence operator reduces to the equal-time product $h_t^{\prime}h_t^{\prime\prime}$, which can be expanded as
\begin{align}
h_t^{\prime} h_{t}^{\prime \prime} &= \frac{1}{32}\left[e^{i\left(\phi_A+\phi_B+2 \phi_p\right)} a_{t-3\tau_m} a_{t-3\tau_m}+e^{i\left(\phi_A+\phi_B+\phi_p\right)} a_{t-3\tau_m} a_{t-\tau_m}\right. +e^{i\left(\phi_A+2 \phi_p\right)} a_{t-3\tau_m} a_{t-2\tau_m} \notag \\
&\quad +e^{i\left(\phi_A+\phi_p\right)} a_{t-3\tau_m} a_{t}+e^{i\left(\phi_A+\phi_B+\phi_p\right)} a_{t-\tau_m} a_{t-3\tau_m}+e^{i\left(\phi_A+\phi_{B}\right)} a_{t-\tau_m} a_{t-\tau_m} +e^{i\left(\phi_A+\phi_p\right)} a_{t-\tau_m} a_{t-2\tau_m} \notag \\
&\quad +e^{i \phi_A} a_{t-\tau_m} a_{t}+e^{i\left(\phi_B+2 \phi_p\right)} a_{t-2\tau_m} a_{t-3\tau_m}+e^{i\left(\phi_B+\phi_p\right)} a_{t-2\tau_m} a_{t-\tau_m} \notag \\
&\quad +e^{2 i \phi_p} a_{t-2\tau_m} a_{t-2\tau_m}+e^{i \phi_p} a_{t-2\tau_m} a_{t} \left.+e^{i\left(\phi_B+\phi_p\right)} a_t a_{t-3\tau_m}+e^{i \phi_B} a_t a_{t-\tau_m}+e^{i \phi_p} a_t a_{t-2\tau_m}+a_t a_{t}\right].
\end{align}

We take the tensor-product input state defined in Eq.~\eqref{eq:inputstate} as the input at port $a$:
\begin{align}
|\psi_{in}\rangle_{\Delta t = 0} &= |\psi_{t-3\tau_m}\rangle |\psi_{t-2\tau_m}\rangle |\psi_{t-\tau_m}\rangle |\psi_{t}\rangle \notag \\
&= \left(\sqrt{p_0}|0\rangle + \sqrt{p_1}|1\rangle\right)_{t-3\tau_m}
   \left(\sqrt{p_0}|0\rangle + \sqrt{p_1}|1\rangle\right)_{t-2\tau_m}
   \left(\sqrt{p_0}|0\rangle + \sqrt{p_1}|1\rangle\right)_{t-\tau_m}
   \left(\sqrt{p_0}|0\rangle + \sqrt{p_1}|1\rangle\right)_{t} \notag \\
&= p_0^2 |0000\rangle + p_0 \sqrt{p_0 p_1} |0100\rangle + p_0 \sqrt{p_0 p_1} |1000\rangle + p_0 p_1 |1100\rangle \notag \\
&\quad + p_0 \sqrt{p_0 p_1} |0010\rangle + p_0 p_1 |0110\rangle + p_0 p_1 |1010\rangle + p_1 \sqrt{p_0 p_1} |1110\rangle \notag \\
&\quad + p_0 \sqrt{p_0 p_1} |0001\rangle + p_0 p_1 |0101\rangle + p_0 p_1 |1001\rangle + p_1 \sqrt{p_0 p_1} |1101\rangle \notag \\
&\quad + p_0 p_1 |0011\rangle + p_1 \sqrt{p_0 p_1} |0111\rangle + p_1 \sqrt{p_0 p_1} |1011\rangle + p_1^2 |1111\rangle.
\end{align}

Then, the coincidence probability in this case can be determined as:
\begin{align}
\mathcal{C}{(0)} &= {\vert h_t^{\prime} h_{t}^{\prime \prime}|\psi_{in}\rangle_{\Delta t} \vert}^2 \notag \\
&= \frac{1}{1024} \left\vert [ 2 e^{i \phi_p} (1-p_1) p_1 + 2 e^{i (\phi_A+\phi_B+\phi_p)} (1-p_1) p_1 + \left(e^{i \phi_A} + e^{i \phi_B}\right) (1-p_1) p_1 \right. \notag \\
&\quad + 2\left(e^{i (\phi_A+\phi_p)} + e^{i (\phi_B+\phi_p)}\right) (1-p_1) p_1 + \left(e^{i (\phi_A+2 \phi_p)} + e^{i (\phi_B+2 \phi_p)}\right) (1-p_1) p_1] |0000\rangle \notag \\
&\quad + [ 2 e^{i (\phi_A+\phi_B+\phi_p)} \sqrt{1-p_1} p_1^{3/2} + \left(e^{i (\phi_A+\phi_p)} + e^{i (\phi_B+\phi_p)}\right) \sqrt{1-p_1} p_1^{3/2} \notag \\
&\quad + \left(e^{i (\phi_A+2 \phi_p)} + e^{i (\phi_B+2 \phi_p)}\right) \sqrt{1-p_1} p_1^{3/2}]|0001\rangle \notag \\
&\quad + [ 2 e^{i (\phi_A+\phi_B+\phi_p)} \sqrt{1-p_1} p_1^{3/2} + \left(e^{i \phi_A} + e^{i \phi_B}\right) \sqrt{1-p_1} p_1^{3/2} \notag \\
&\quad + \left(e^{i (\phi_A+\phi_p)} + e^{i (\phi_B+\phi_p)}\right) \sqrt{1-p_1} p_1^{3/2} ]|0100\rangle + [ 2 e^{i (\phi_A+\phi_B+\phi_p)} p_1^2 ]|0101\rangle \notag \\
&\quad + [ 2 e^{i \phi_p} \sqrt{1-p_1} p_1^{3/2} + \left(e^{i (\phi_A+\phi_p)} + e^{i (\phi_B+\phi_p)}\right) \sqrt{1-p_1} p_1^{3/2}  \notag \\
&+ \left(e^{i (\phi_A+2 \phi_p)} + e^{i (\phi_B+2 \phi_p)}\right) \sqrt{1-p_1} p_1^{3/2}]|0010\rangle \notag \\
&\quad + [ \left(e^{i (\phi_A+2 \phi_p)} + e^{i (\phi_B+2 \phi_p)}\right) p_1^2]|0011\rangle  + [ \left(e^{i (\phi_A+\phi_p)} + e^{i (\phi_B+\phi_p)}\right) p_1^2]|0110\rangle \notag \\
&\quad + [ 2 e^{i \phi_p} \sqrt{1-p_1} p_1^{3/2} + \left(e^{i \phi_A} + e^{i \phi_B}\right) \sqrt{1-p_1} p_1^{3/2} + \left(e^{i (\phi_A+\phi_p)} + e^{i (\phi_B+\phi_p)}\right) \sqrt{1-p_1} p_1^{3/2}]|1000\rangle \notag \\
&\quad + [ \left(e^{i (\phi_A+\phi_p)} + e^{i (\phi_B+\phi_p)}\right) p_1^2]|1001\rangle + [ \left(e^{i \phi_A} + e^{i \phi_B}\right) p_1^2]|1100\rangle  + \left. [ 2 e^{i \phi_p} p_1^2 ]|1010\rangle \right\vert^2.
\label{Eq.2}
\end{align}

According to Eq.~\ref{Eq.2}, the quantity $\mathcal{C}(0)$ is obtained by taking the modulus squared of the total probability amplitude, which is mathematically equivalent to summing the squared magnitudes of the coefficients associated with each basis state in the expanded form of $h_t^{\prime} h_t^{\prime\prime} |\psi_{\mathrm{in}}\rangle$.  
Because the full expansion is lengthy, containing contributions from all allowed temporal-path alternatives, we show only a subset of terms below for brevity:
\begin{align}
\mathcal{C}{(0)} 
&=  
\frac{1}{1024}(20 p_1^2 - 40 p_1^3 + 36 p_1^4 
+ 32(1-p_1)p_1^3
+ 16 p_1^2 \cos\phi_A + 8p_1^2 \cos(\phi_B+\phi_p) \notag \\
&\quad
+ 16 p_1^2 \cos\phi_B - 32 p_1^3 \cos\phi_B + 16 p_1^4 \cos\phi_B + 8p_1^2 \cos(\phi_A-\phi_p) \notag \\
&\quad
+ 8 p_1^2 \cos(\phi_A+\phi_B) - 16 p_1^3 \cos(\phi_A+\phi_B) + 8 p_1^4 \cos(\phi_A+\phi_B)
+ \cdots).
\end{align}

With $\phi_p = \pi$, we have
\begin{align}
\mathcal{C}{(0)} = \frac{p_1^2}{128}  [(1+p_1^2+p_1^2 cos(\phi_A-\phi_B)+p_0^2 cos(\phi_A+\phi_B))].
\end{align}

In the low-excitation regime, where $p_1 \ll p_0$, the above expression simplifies to 
\begin{align}
\mathcal{C}{(0)} = \frac{ p_1^2}{128} [1+p_0^2 cos(\phi_A+\phi_B)].
\end{align}

\subsubsection{$\Delta t = \pm \tau_p$}

For the characteristic delays $\Delta t = \pm \tau_p$,  
we focus on the case $\Delta t = \tau_p$, as the analysis for $\Delta t = -\tau_p$ proceeds analogously. The corresponding field operator governing the coincidence process can be written as

%%equation with better alignment
\begin{align}
h_t^{\prime} h_{t+\tau_p}^{\prime\prime}
&= \frac{1}{32}\Big[
e^{i(\phi_A+\phi_B+2\phi_p)} a_{t-3\tau_m} a_{t-\tau_m}
+ e^{i(\phi_A+\phi_B+\phi_p)} a_{t-3\tau_m} a_{t+\tau_m}
\notag\\
&\quad
+ e^{i(\phi_A+2\phi_p)} a_{t-3\tau_m} a_t
+ e^{i(\phi_A+\phi_p)} a_{t-3\tau_m} a_{t+2\tau_m}
\notag\\
&\quad
+ e^{i(\phi_A+\phi_B+\phi_p)} a_{t-\tau_m} a_{t-\tau_m}
+ e^{i(\phi_A+\phi_B)} a_{t-\tau_m} a_{t+\tau_m}
\notag\\
&\quad
+ e^{i(\phi_A+\phi_p)} a_{t-\tau_m} a_t
+ e^{i\phi_A} a_{t-\tau_m} a_{t+2\tau_m}
\notag\\
&\quad
+ e^{i(\phi_B+2\phi_p)} a_{t-2\tau_m} a_{t-\tau_m}
+ e^{i(\phi_B+\phi_p)} a_{t-2\tau_m} a_{t+\tau_m}
\notag\\
&\quad
+ e^{2i\phi_p} a_{t-2\tau_m} a_t
+ e^{i\phi_p} a_{t-2\tau_m} a_{t+2\tau_m}
\notag\\
&\quad
+ e^{i(\phi_B+\phi_p)} a_t a_{t-\tau_m}
+ e^{i\phi_B} a_t a_{t+\tau_m}
+ e^{i\phi_p} a_t a_t
+ a_t a_{t+2\tau_m}
\Big].
\end{align}

The corresponding input state at port $a$ can be expressed as a tensor product of six temporal modes,

\begin{align}
|\psi_{in}\rangle_{\Delta t=\tau_p}
&= |\psi_{t-3\tau_m}\rangle |\psi_{t-2\tau_m}\rangle
   |\psi_{t-\tau_m}\rangle |\psi_t\rangle
   |\psi_{t+\tau_m}\rangle |\psi_{t+2\tau_m}\rangle
\notag\\
&= \sum_{b\in\{0,1\}^6}
   p_0^{\frac{6-\mathrm{wt}(b)}{2}}
   p_1^{\frac{\mathrm{wt}(b)}{2}}
   |b\rangle ,
\end{align}
\noindent
where $\mathrm{wt}(b)$ denotes the number of ones in the six-bit string $b$.

Under the experimental setting $\phi_p = \pi$, the coincidence probability at $\Delta t = \pm\tau_p$ reads:
\begin{align}
    \mathcal{C}{(\pm \tau_p)} &= \left\vert h_t^{\prime} h_{t+\Delta t}^{\prime \prime} \vert \psi_{in} \rangle_{\Delta t = 2\tau_m} \right\vert^2
    \xrightarrow{\phi_p = \pi} \frac{p_1^2}{512}  \left[1 + 2p_1 + 4p_1^2 + p_1^2 \cos(\phi_A - \phi_B) + p_0^2 \cos(\phi_A + \phi_B)\right]. 
\end{align}

In the weak-excitation limit $p_1 \ll 1$, 
this expression reduces to

\begin{align}
    \mathcal{C}{(\pm \tau_p)}= \frac{p_1^2}{512}  \left[1 + p_0^2\cos(\phi_A + \phi_B)\right] 
\end{align}

%\noindent
\subsubsection{$\Delta t = \pm \tau_m$}

Because $\tau_p=2\tau_m$ in our experimental setup, the temporal modes at delays $\tau_m$ and $\tau_p-\tau_m$ become degenerate and thus merge.
Consequently, coincidence events at the characteristic delays $\Delta t = \pm \tau_m$ involve five distinct temporal modes at port $a$. Without loss of generality, we consider the case $\Delta t = +\tau_m$, since the analysis for $\Delta t = -\tau_m$ is analogous. The corresponding field operator can be written as:

\begin{align}
h_t^{\prime} h_{t+\tau_m}^{\prime\prime}
&= \frac{1}{32}\Big[
e^{i(\phi_A+\phi_B+2\phi_p)} a_{t-3\tau_m} a_{t-2\tau_m}
+ e^{i(\phi_A+\phi_B+\phi_p)} a_{t-3\tau_m} a_t
\notag\\
&\quad
+ e^{i(\phi_A+2\phi_p)} a_{t-3\tau_m} a_{t-\tau_m}
+ e^{i(\phi_A+\phi_p)} a_{t-3\tau_m} a_{t+\tau_m}
\notag\\
&\quad
+ e^{i(\phi_A+\phi_B+\phi_p)} a_{t-\tau_m} a_{t-2\tau_m}
+ e^{i(\phi_A+\phi_B)} a_{t-\tau_m} a_t
\notag\\
&\quad
+ e^{i(\phi_A+\phi_p)} a_{t-\tau_m} a_{t-\tau_m}
+ e^{i\phi_A} a_{t-\tau_m} a_{t+\tau_m}
\notag\\
&\quad
+ e^{i(\phi_B+2\phi_p)} a_{t-2\tau_m} a_{t-2\tau_m}
+ e^{i(\phi_B+\phi_p)} a_{t-2\tau_m} a_t
\notag\\
&\quad
+ e^{2i\phi_p} a_{t-2\tau_m} a_{t-\tau_m}
+ e^{i\phi_p} a_{t-2\tau_m} a_{t+\tau_m}
\notag\\
&\quad
+ e^{i(\phi_B+\phi_p)} a_t a_{t-2\tau_m}
+ e^{i\phi_B} a_t a_t
\notag\\
&\quad
+ e^{i\phi_p} a_t a_{t-\tau_m}
+ a_t a_{t+\tau_m}
\Big].
\end{align}
The associated input quantum state at port $a$ can be expressed as a tensor product of five time modes,
\begin{align}
|\psi_{in}\rangle_{\Delta t=\tau_m}
&= |\psi_{t-3\tau_m}\rangle |\psi_{t-2\tau_m}\rangle
   |\psi_{t-\tau_m}\rangle |\psi_t\rangle |\psi_{t+\tau_m}\rangle
\notag\\
&= \left(\sqrt{p_0}|0\rangle+\sqrt{p_1}|1\rangle\right)_{t-3\tau_m}
   \left(\sqrt{p_0}|0\rangle+\sqrt{p_1}|1\rangle\right)_{t-2\tau_m}
   \left(\sqrt{p_0}|0\rangle+\sqrt{p_1}|1\rangle\right)_{t-\tau_m}
\notag\\
&\quad
   \left(\sqrt{p_0}|0\rangle+\sqrt{p_1}|1\rangle\right)_t
   \left(\sqrt{p_0}|0\rangle+\sqrt{p_1}|1\rangle\right)_{t+\tau_m}
\notag\\
&= p_0^{5/2}|00000\rangle + p_0^2 p_1^{1/2}|00001\rangle
 + p_0^2 p_1^{1/2}|00010\rangle + p_0^{3/2}p_1|00011\rangle
\notag\\
&\quad
 + p_0^2 p_1^{1/2}|00100\rangle + p_0^{3/2}p_1|00101\rangle
 + p_0^{3/2}p_1|00110\rangle + p_0 p_1^{3/2}|00111\rangle
\notag\\
&\quad
 + p_0^2 p_1^{1/2}|01000\rangle + p_0^{3/2}p_1|01001\rangle
 + p_0^{3/2}p_1|01010\rangle + p_0 p_1^{3/2}|01011\rangle
\notag\\
&\quad
 + p_0^{3/2}p_1|01100\rangle + p_0 p_1^{3/2}|01101\rangle
 + p_0 p_1^{3/2}|01110\rangle + p_0^{1/2}p_1^2|01111\rangle
\notag\\
&\quad
 + p_0^2 p_1^{1/2}|10000\rangle + p_0^{3/2}p_1|10001\rangle
 + p_0^{3/2}p_1|10010\rangle + p_0 p_1^{3/2}|10011\rangle
\notag\\
&\quad
 + p_0^{3/2}p_1|10100\rangle + p_0 p_1^{3/2}|10101\rangle
 + p_0 p_1^{3/2}|10110\rangle + p_0^{1/2}p_1^2|10111\rangle
\notag\\
&\quad
 + p_0^{3/2}p_1|11000\rangle + p_0 p_1^{3/2}|11001\rangle
 + p_0 p_1^{3/2}|11010\rangle + p_0^{1/2}p_1^2|11011\rangle
\notag\\
&\quad
 + p_0 p_1^{3/2}|11100\rangle + p_0^{1/2}p_1^2|11101\rangle
 + p_0^{1/2}p_1^2|11110\rangle + p_1^{5/2}|11111\rangle .
\end{align}

With $\phi_p=\pi$, the coincidence probability at $\Delta t=\pm\tau_m$ is
\begin{equation}
\begin{aligned}
\mathcal{C}{(\pm \tau_m)} = \left\vert h_t^{\prime} h_{t+\Delta t}^{\prime \prime} \vert \psi_{\mathrm{in}} \rangle_{\Delta t =\tau_m} \right\vert^2
\xrightarrow{\phi_p = \pi} \frac{p_1^2}{1024} [ & 5 + 2p_1 + 8p_1^2 - 4 \cos(\phi_A - \phi_B) \\
& + 8p_1 \cos(\phi_A - \phi_B) - 4p_1^2 \cos(\phi_A - \phi_B) \\
& - 4p_1^2 \cos(\phi_A + \phi_B)].
\end{aligned}
\end{equation}

In the weak-excitation limit $p_1\to 0$,
the above expression can be approximated as
\begin{align}
\mathcal{C}(\pm \tau_m)=\frac{p_1^2}{1024}\Big[5-4p_0^2\cos(\phi_A-\phi_B)\Big].
\end{align}

\subsubsection{$\Delta t = \pm (\tau_p+\tau_m)$}

For the characteristic delays $\Delta t=\pm(\tau_p+\tau_m)$, i.e., $\Delta t=\pm 3\tau_m$, the field operator relevant to the coincidence process is given by the following expression. For clarity, we present the explicit expression for the $+3\tau_m$ side peak; the $-3\tau_m$ case follows analogously (equivalently, by the symmetry $\Delta t\to-\Delta t$):
\begin{align}
h_t^{\prime} h_{t+3\tau_m}^{\prime\prime}
&= \frac{1}{32}\Big[
e^{i(\phi_A+\phi_B+2\phi_p)} a_{t-3\tau_m} a_t
+ e^{i(\phi_A+\phi_B+\phi_p)} a_{t-3\tau_m} a_{t+2\tau_m}
\notag\\
&\quad
+ e^{i(\phi_A+2\phi_p)} a_{t-3\tau_m} a_{t+\tau_m}
+ e^{i(\phi_A+\phi_p)} a_{t-3\tau_m} a_{t+3\tau_m}
\notag\\
&\quad
+ e^{i(\phi_A+\phi_B+\phi_p)} a_{t-\tau_m} a_t
+ e^{i(\phi_A+\phi_B)} a_{t-\tau_m} a_{t+2\tau_m}
\notag\\
&\quad
+ e^{i(\phi_A+\phi_p)} a_{t-\tau_m} a_{t+\tau_m}
+ e^{i\phi_A} a_{t-\tau_m} a_{t+3\tau_m}
\notag\\
&\quad
+ e^{i(\phi_B+2\phi_p)} a_{t-2\tau_m} a_t
+ e^{i(\phi_B+\phi_p)} a_{t-2\tau_m} a_{t+2\tau_m}
\notag\\
&\quad
+ e^{2i\phi_p} a_{t-2\tau_m} a_{t+\tau_m}
+ e^{i\phi_p} a_{t-2\tau_m} a_{t+3\tau_m}
\notag\\
&\quad
+ e^{i(\phi_B+\phi_p)} a_t a_t
+ e^{i\phi_B} a_t a_{t+2\tau_m}
\notag\\
&\quad
+ e^{i\phi_p} a_t a_{t+\tau_m}
+ a_t a_{t+3\tau_m}
\Big].
\end{align}

The associated input quantum state at port $a$ can be expressed as a tensor product of seven temporal modes,
\begin{align}
|\psi_{in}\rangle_{\Delta t=3\tau_m}
&= p_0^{7/2}|0000000\rangle
+ p_0^3p_1^{1/2} \sum_{\mathrm{wt}(b)=1}|b\rangle
+ p_0^{5/2}p_1 \sum_{\mathrm{wt}(b)=2}|b\rangle
\notag\\
&\quad
+ p_0^2p_1^{3/2} \sum_{\mathrm{wt}(b)=3}|b\rangle
+ p_0^{3/2}p_1^2 \sum_{\mathrm{wt}(b)=4}|b\rangle
+ p_0p_1^{5/2} \sum_{\mathrm{wt}(b)=5}|b\rangle
\notag\\
&\quad
+ p_0^{1/2}p_1^3 \sum_{\mathrm{wt}(b)=6}|b\rangle
+ p_1^{7/2}|1111111\rangle .
\end{align}
\noindent
Here, $\mathrm{wt}(b)$ denotes the Hamming weight of the bit string $b$, i.e., the number of ones in $b$.

We further present the analytical expression for the coincidence probability associated with the side peaks $\Delta t=\pm(\tau_m+\tau_p)$ under the phase condition $\phi_p=\pi$:
\begin{align}
    \mathcal{C}(\pm (\tau_m+\tau_p)) &= \left\vert h_t^{\prime} h_{t+\Delta t}^{\prime \prime} \vert \psi_{in} \rangle_{\Delta t = 3\tau_m} \right\vert^2
    \xrightarrow{\phi_p = \pi} \frac{p_1^2}{1024} \left[12p_1^2+2p_1+1\right]. 
\end{align}

We note that the coincidence probability at $\Delta t=\pm(\tau_p+\tau_m)$ does not depend on the phase setting of either analysis AMZI.

\subsection{The CHSH Bell parameter evaluation}

To evaluate the Bell parameters theoretically, it is sufficient to specify the single-photon probability $p_1$, which determines the coincidence probabilities entering the correlation functions at $\Delta t=0$ and $\Delta t=\pm\tau_p$. Using the measurement settings implemented in the experiment,
$(\phi_A,\phi_B,\phi'_A,\phi'_B)=(0,\pi/4,\pi/2,3\pi/4)$,
we substitute these phases into the expressions for the correlation functions to obtain
$E(\phi_A,\phi_B)$, $E(\phi_A,\phi'_B)$, $E(\phi'_A,\phi_B)$, and $E(\phi'_A,\phi'_B)$.
Combining them in the standard CHSH form~\cite{Clauser1969PRL} yields
\begin{align}
S(0)= 2\sqrt{2}\frac{(1-p_1)^2}{p_1^2+1}, 
\qquad
S(\pm\tau_p)= 2\sqrt{2}\frac{(1-p_1)^2}{1+2p_1+4p_1^2}.
\label{V}
\end{align}

These expressions show that the Bell parameters for the $\ket{2}$ and $\ket{11}$ states are controlled by the single-photon probability $p_1$. In the ideal limit $p_1\to 0$, both $S(0)$ and $S(\pm\tau_p)$ approach the maximal quantum value $2\sqrt{2}$, corresponding to near-perfect two-photon interference. As $p_1$ increases, higher-order contributions reduce the correlation visibility and hence decrease $S$. For sufficiently large $p_1$, both expressions drop below the classical bound of $2$, signaling the loss of Bell violation. This monotonic dependence underscores the importance of suppressing the single-photon component to maximize nonclassical correlations.

\section{Further experimental details}

\subsection{Phase-locking scheme for the AMZIs}
\begin{figure}[htb]
\centering 
\includegraphics[width=0.9\textwidth]{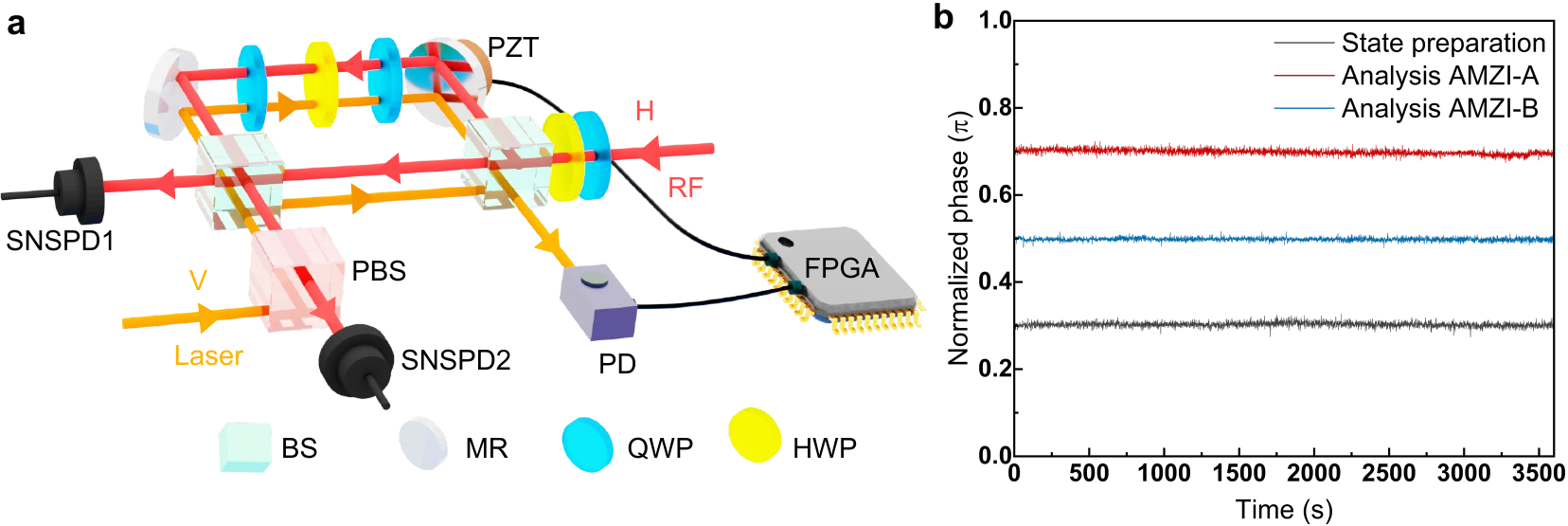}
\caption{\textbf{AMZI phase-locking setup and long-term phase stability.}
\textbf{a}, Schematic of the active phase-stabilization system based on a photodetector (PD), a field-programmable gate array (FPGA)-based proportional--integral--derivative (PID) controller, and a piezoelectric transducer (PZT)-mounted mirror (MR).
\textbf{b}, Measured phase fluctuations over one hour for three AMZIs.}
\label{S5}
\end{figure}

All three AMZIs used in the experiment are actively phase locked using an identical scheme, as illustrated in Fig.~\ref{S5}(a). The locking starts with a vertically polarized laser beam, orthogonal to the RF signal, whose wavelength and intensity are actively stabilized. The laser beam is injected through an output port of each interferometer and counter-propagates alongside the horizontally polarized RF signal. The locking and signal beams are vertically separated by $\sim 6$~mm, so the locking beam follows the same interferometric path while minimizing contamination the RF  signal. A photodiode (PD) monitors the locking-beam interference output intensity, and an FPGA-based PID controller compares it to a preset reference to generate an error signal. The resulting feedback voltage is amplified to drive a PZT-mounted mirror. With a loop update time of $100~\mu$s, the feedback is able to stabilize the interferometer at any user-selected phase setpoint.

For target phases of $0$ or $\pi$, the interferometer output intensity has vanishing slope with respect to phase so direct locking to these operating points is difficult. 
To circumvent this problem, we insert a QWP--HWP--QWP set (all at $45^{\circ}$) into the long arm of each AMZI. This wave-plate set provides an adjustable relative phase between the H- and V-polarized components while returning each polarization to itself, thereby preserving the signal polarization. We then lock the counter-propagating reference beam at an intermediate phase (e.g., $\pi/2$), where the interference intensity varies approximately linearly with phase, and subsequently rotate the HWP to shift the signal phase to the desired value (including $0$ or $\pi$).
This scheme ensures stable preparation of the relevant two-photon states and enables flexible phase-projection measurements in Alice's and Bob's  
analysis interferometers.

We evaluate the locking stability by continuously phase locking the three AMZIs for one hour using a continuous-wave laser. As shown in Fig.~\ref{S5}(b), the measured phase fluctuations have a standard deviation of $\sim 0.006\pi$, satisfying the phase-stability requirements of the experiment.

\subsection{Phase and excitation-power dependence of the AMZI-prepared quantum state}

\begin{figure}[hbt]
\centering 
\includegraphics[width=0.9\textwidth]{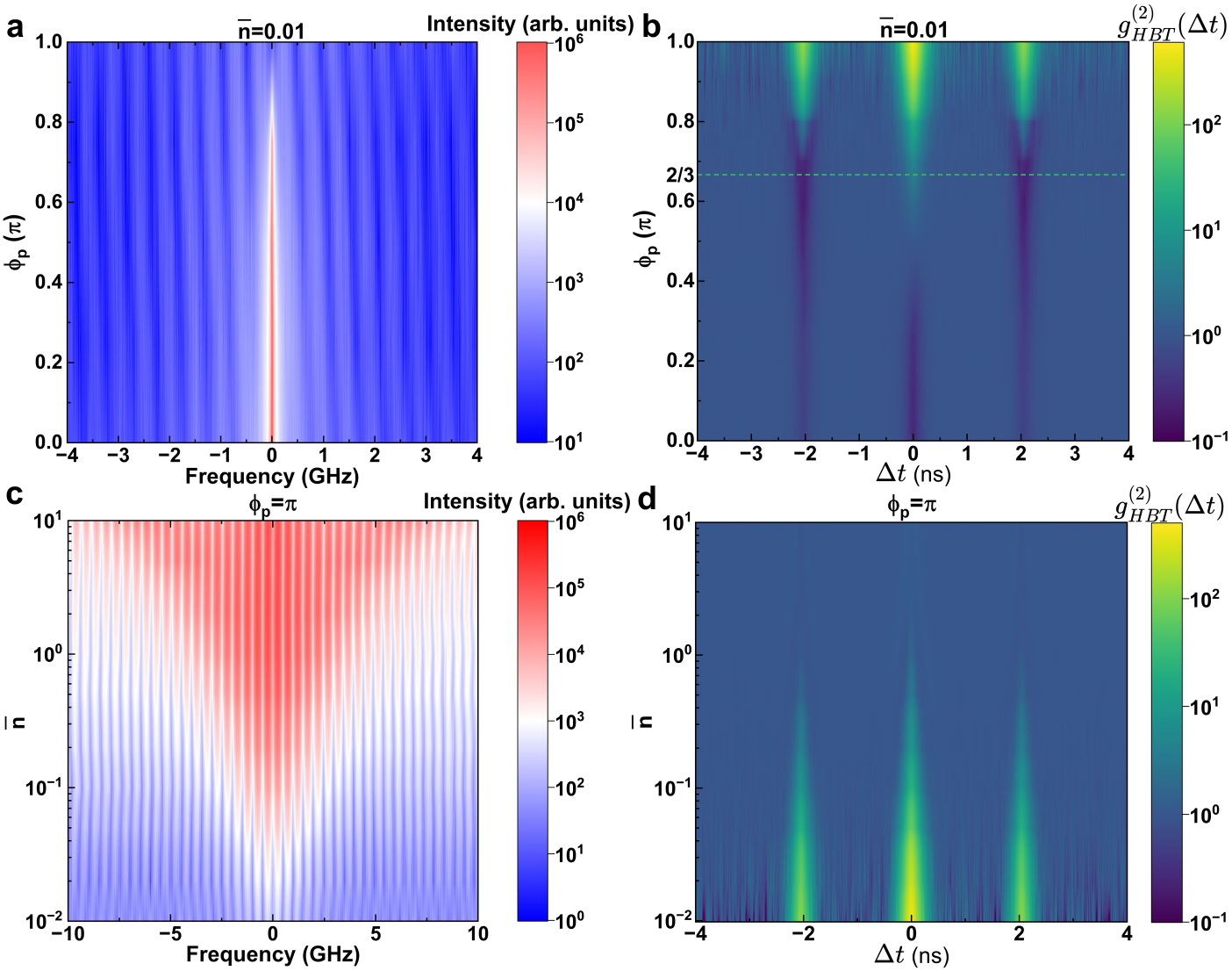}
\caption{\textbf{Phase- and power-dependent response of the state-preparation AMZI.}
\textbf{a,b}, High-resolution RF spectra (\textbf{a}) and corresponding two-dimensional maps of the second-order correlation $g^{(2)}_{HBT}(\Delta t)$ (\textbf{b}) measured as a function of the AMZI phase $\phi_p$ at $\bar{n}=0.01$. The dashed line at $\phi_p=\frac{2\pi}{3}$ indicates the minimum of $g^{(2)}_{HBT}(\pm\tau_p)$, caused by destructive interference of the  $\ket{11}$ component in the AMZI-prepared field.
\textbf{c,d}, Power dependence measured at fixed phase $\phi_p=\pi$: high-resolution RF spectra (\textbf{c}) and corresponding $g^{(2)}_{HBT}(\Delta t)$ maps (\textbf{d}) recorded as a function of the excitation power $\bar{n}$.}
\label{S6}
\end{figure}

Acting as a phase-tunable filter, the state-preparation AMZI plays a vital role in shaping the RF field and generating the desired two-photon states. Figure~\ref{S6}(a) shows the high-resolution spectra of the RF after passing through the state-preparation AMZI under different phase settings. The interference fringes shift along the frequency axis with varying $\phi_p$, and after a full $\pi$ phase shift, the fringes translate by one complete period ($\sim467$ MHz), corresponding to the delay imbalance $\tau_p$  
between the long and short arms of the AMZI.  
Notably, when the interferometer phase is set to $\pi$, the laser-like component (0~GHz) undergoes destructive interference and reaches its minimum intensity. In contrast, at $\phi_p=0$ the laser-like component is maximized,  with no intrinsic attenuation by the AMZI, while for $0<\phi_p<\pi$ the degree of attenuation varies smoothly with $\phi_p$.

Figure~\ref{S6}(b) presents the corresponding second-order autocorrelation functions measured under When the interferometer phase is set to $\phi_p = \pi$.  We observe pronounced photon bunching, with $g^{(2)}_{HBT}(0)\approx 511$. This behavior is consistent with Fig.~\ref{S6}(a) and reflects near-maximal destructive interference of the vacuum--one-photon superposition, which suppresses the single-photon component at the AMZI output and thereby enhances the two-photon correlations.
When $\phi_p$ is tuned from $\pi$ to $0$, the weight of the single photon component increases continuously, driving a smooth crossover from photon bunching to antibunching; at $\phi_p=0$, we obtain $g^{(2)}_{HBT}(0)\approx 0.27$.

In Fig.~\ref{S6}(b), the two vertical features at $\Delta t=\pm\tau_p$ originate from coincidence events in which the two photons occupy adjacent time bins, corresponding to the time-bin-separated two-photon state $\ket{11}$. As the state-preparation AMZI phase $\phi_p$ is tuned, the intensity of these $\pm\tau_p$ features varies markedly, revealing phase-dependent two-photon correlation within the $\ket{11}$ sector.
In particular, $g^{(2)}_{HBT}(\pm\tau_p)$ reaches a clear minimum at $\phi_p=2\pi/3$ (minimum value $g^{(2)}_{HBT}(\pm\tau_p)\simeq 0.17$ extracted from the data), indicating destructive interference of the relevant two-photon amplitude.
This behavior is consistent with the analytical output state in Eq.~\eqref{out.1}, where the $\ket{1_d1_d}$ contribution (fourth line) carries the phase-dependent factor $(1+e^{i\phi_p}+e^{2i\phi_p})$, which vanishes at $\phi_p=2\pi/3$. By comparison, the correlation at $\Delta t=0$ is governed primarily by the phase-dependent suppression of the single-photon component at the AMZI output, while the amplitude of $\ket{2}$ is phase-independent. At $\phi_p=0$ the suppression vanishes, yielding the smallest value of $g^{(2)}_{HBT}(0) = 0.27$, whereas at $\phi_p=\pi$ the single-photon component is most strongly suppressed, leading to the largest superbunching of $g^{(2)}_{HBT}(0) = 511$. Intermediate phases produce a continuous evolution between these two limits.

Figures~\ref{S6}(c) and~\ref{S6}(d) show the excitation-flux dependence of the high-resolution spectra and the second-order autocorrelation at the output of the state-preparation AMZI. Even at $\bar{n}=10$, the spectrally sharp laser-like component near 0~GHz remains strongly suppressed, demonstrating the robustness of the two-path interference in rejecting the first-order coherent fraction. With increasing excitation, the main change is therefore not the filtering efficiency, but the internal composition of the RF field itself. In particular, the coherent single-photon amplitude in the AMZI output scales as $\sqrt{p_1}p_0$ (second line of Eq.~\eqref{out.1}) and decreases as $p_0$ is reduced, causing spectral weight to shift toward the broadband component and thus lowering the overall first-order coherence.

In parallel, the prepared quantum state progressively loses the interference-induced structure responsible for pronounced bunching. Accordingly, the zero-delay correlation $g^{(2)}_{HBT}(0)$ decreases rapidly and approaches unity at $\bar{n}\approx 2$, indicating the disappearance of bunching. As established in our previous work~\cite{wang2025coherence}, the relation $g^{(2)}_{HBT}(0)=1/p_1^2$ quantitatively captures this trend by linking the bunching strength to the population of the single-quanta sector. The same mechanism governs $g^{(2)}_{HBT}(\pm\tau_p)=\frac{1-2p_1}{4p_1^2}$, which exhibits a similar excitation-flux dependence.

\section {Methods for simulating the experimental data}

\subsection{Extraction of $p_1$ from autocorrelation data}

\begin{figure}[htb]
\centering 
\includegraphics[width=0.45\textwidth]{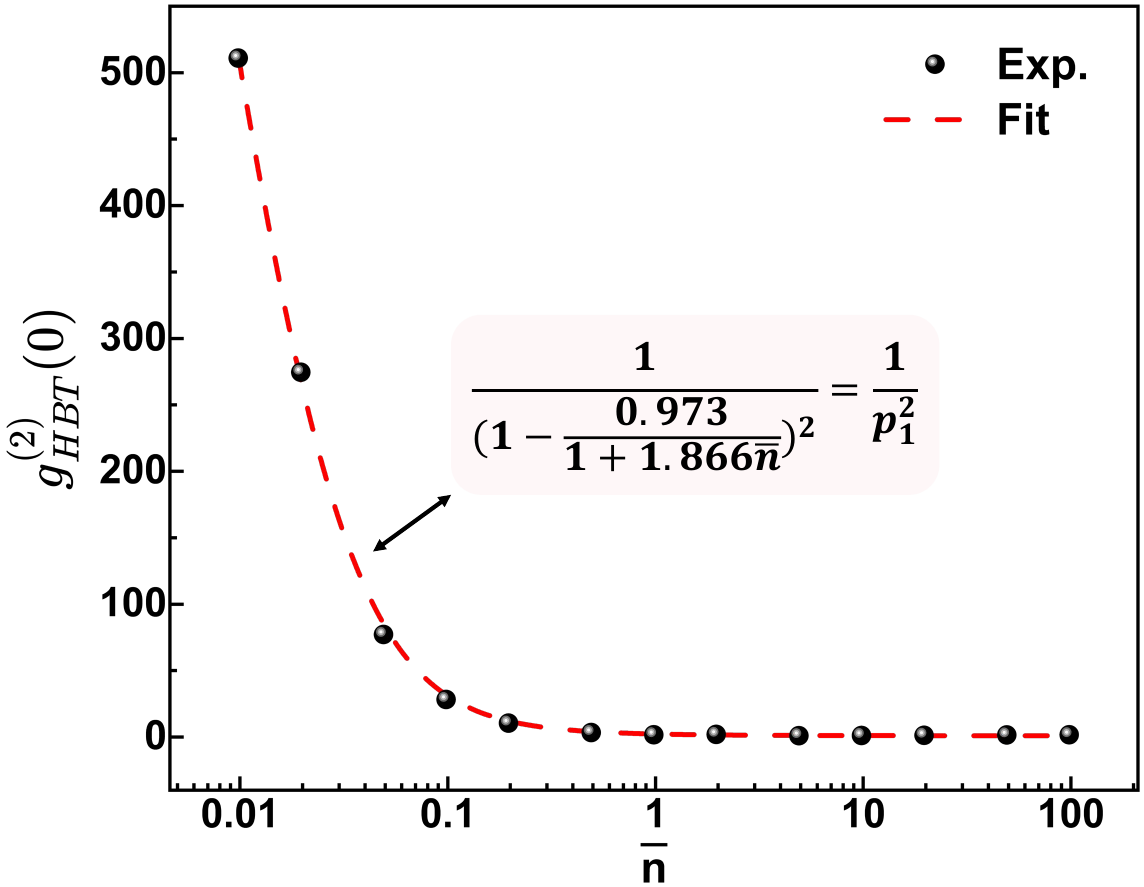}
\caption{\textbf{Zero-delay autocorrelation versus excitation power for the AMZI-prepared field.} 
}
\label{S4}
\end{figure}

In this work, the excitation-power-dependent single-photon probability $p_1$ is the key input parameter for simulating the coincidence probabilities and the resulting Bell parameter $S$. We extract $p_1$ from the measured autocorrelation of the state-preparation AMZI output. When the AMZI phase is set to $\phi_p=\pi$, the output contains a two-photon Fock-state component $\ket{2}$; because of this component, the zero-delay correlation of the AMZI-filtered RF obeys~\cite{wang2025coherence}
\begin{align}
g^{(2)}_{HBT}(0)=\frac{1}{p_1^2}.
\end{align}
For completeness, the side-peak correlation at delay $\tau_p$ is given by $g^{(2)}_{HBT}(\pm\tau_p)=\frac{1+2p_1}{4p_1^2}$~\cite{wang2025coherence}. 
Therefore, measuring $g^{(2)}_{HBT}(0)$ as a function of excitation power directly determines $p_1(\bar{n})$.

To parameterize the excitation-power dependence, we start from the relation (neglecting pure dephasing)~\cite{wang2025coherence,HuangGQ2026}
\begin{align}
p_1 = 1-\frac{1}{1+2\bar{n}\eta},
\end{align}
where $\eta$ is the absorption efficiency in the weak-excitation limit and $\bar{n}$ is the mean excitation number. To account for experimental imperfections, we adopt an empirical modification of this functional form and fit the measured $g^{(2)}(0)$ using
\begin{align}
g^{(2)}_{HBT}(0)=\frac{1}{p_1^2}
=\frac{1}{\left(1-\frac{A}{1+B\bar{n}}\right)^2},
\end{align}
where $A$ and $B$ are fitting parameters. The fit (Fig.~\ref{S4}) shows excellent agreement with the experimental data and yields $A=0.973$ and $B=1.866$. Consequently, the extracted single-photon probability used throughout our simulations is
\begin{align}
p_1(\bar{n})=1-\frac{0.973}{1+1.866\bar{n}}.
\end{align}

This calibrated function $p_1(\bar{n})$ is then substituted into the analytical expressions for the coincidence probabilities (Section~I.5) and, in particular, into Eq.~\eqref{V} to obtain the excitation-power dependence of the Bell parameter $S$ shown in Fig.~5\textbf{b} of the main text.

\subsection{The concatenation of coincidence probabilities at characteristic time points}

\begin{figure}[htb]
\centering 
\includegraphics[width=0.45\textwidth]{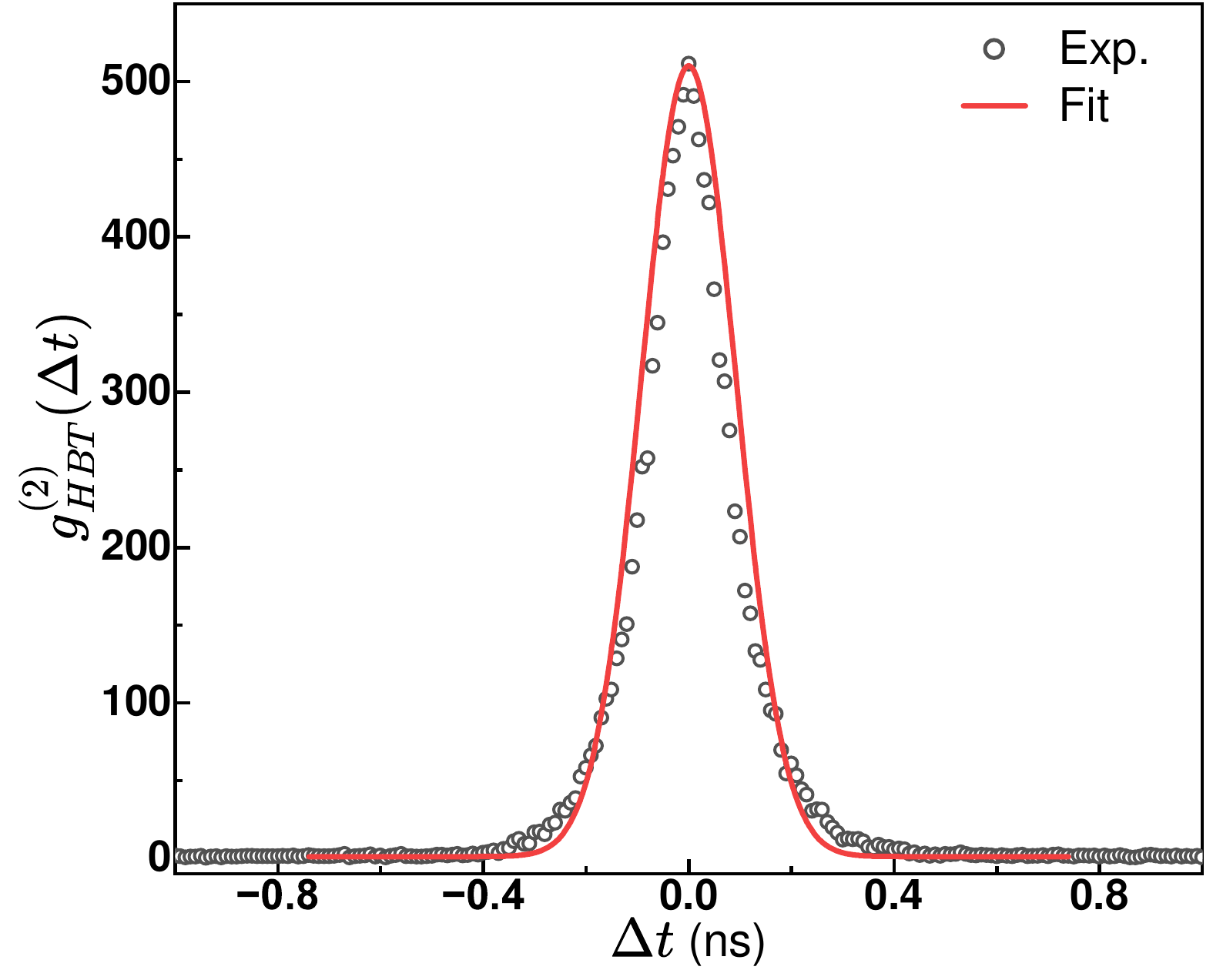}
\caption{\textbf{Autocorrelation at the output of the AMZI-prepared output.}}
\label{S3}
\end{figure}

In Section~I.3, we have rigorously derived the coincidence probabilities at the seven characteristic time delays, namely $\mathcal{C}(0)$, $\mathcal{C}(\pm\tau_p)$, $\mathcal{C}(\pm\tau_m)$, and $\mathcal{C}\!\left(\pm(\tau_p+\tau_m)\right)$. In this section, we briefly discuss how these discrete-delay probabilities are connected by the second-order autocorrelation function of the RF signal after the state-preparation AMZI, thereby enabling the theoretical construction of a two-dimensional correlation map of coincidence probabilities versus phase and time delay.

The second-order autocorrelation function of the two-photon $\ket{2}$ state is expressed as $g^{(2)}_{HBT}(\Delta t) = A e^{-(\Delta  t)^2/T_2^2}$~\cite{dalibard1983correlation}, where $A$ is a fitting parameter and $T_2 = 134.4~\mathrm{ps}$ represents the coherence time of the QD. As shown in Fig.~\ref{S3}, the fitting yields excellent agreement with the experimental data.

Because the separation between neighboring coincidence peaks greatly exceeds the QD coherence time $T_2$ (i.e., $\tau_p,\tau_m\gg T_2$), coincidence events at different characteristic delays are effectively independent. Accordingly, we construct the second-order correlation function of the prepared state as a product of independently normalized peak contributions,
\begin{align}
g^{(2)}_{HBT}(\Delta t)
&=\prod_{j\in\{0,\pm\tau_p,\pm\tau_m,\pm(\tau_p+\tau_m)\}}
\frac{\mathcal{C}_{j}(\Delta t)}{\mathcal{C}_{j}(\infty)} ,
\label{eq:g2_product}
\end{align}
where each factor is normalized such that $\mathcal{C}_{j}(\Delta t)/\mathcal{C}_{j}(\infty)\to 1$ for $|\Delta t|\gg T_2$. This construction is used to generate the simulated correlation maps shown in Figs.~3\textbf{d} and~3\textbf{e} of the main text.

\section{Scheme comparison}

\begin{table}[hbt]
\centering
\caption{Comparison of photon-pair generation platforms and key performance metrics.}
\label{tab:photon_pair_comparison}

\renewcommand{\arraystretch}{1.25}
\setlength{\tabcolsep}{4pt}

\begin{tabular}{@{} c c c c c c c c @{}}
\toprule
\textbf{Reference} & 
\textbf{Platform} & 
\textbf{Source} & 
$g^{(2)}_{HBT}(0)$ & 
\makecell[c]{\textbf{Pair Rate} \\ \textbf{(Hz)}$^a$} & 
\makecell[c]{\textbf{Pump} \\ \textbf{Power}} & 
\makecell[c]{\textbf{Rate/Power} \\ \textbf{(Hz/mW)}$^a$} & 
$\mathcal{V}$ \\
\midrule

\makecell[c]{This work} & 
\makecell[c]{InAs QD} & 
\makecell[c]{RF (Heitler) \\ + AMZI} & 
511 & 
66.07$^b$ & 
31.9 pW & 
$2.1 \times 10^9$$^b$ & 
95\% \\

\makecell[c]{Liu \textit{et al.}\cite{liu2024violation}} & 
\makecell[c]{InAs QD} & 
\makecell[c]{RF (Heitler) \\ + F-P filter} & 
232 & 
0.26 & 
7.2 pW & 
$3.6 \times 10^7$ & 
95\% \\

\makecell[c]{Wang \textit{et al.}\cite{wang2025purcell}} & 
\makecell[c]{$^{87}$Rb atom} & 
\makecell[c]{RF (Heitler) \\ + F-P filter} & 
100 & 
16 & 
$\sim\mu$W$^c$ & 
$\sim10^{4}$$^c$ & 
93\% \\

\makecell[c]{Peiris \textit{et al.}\cite{peiris2017franson}} & 
\makecell[c]{InAs QD} &  
\makecell[c]{RF (Mollow triplet \\ sidebands)} & 
8 & 
$10^{4}$ & 
$\sim\mu$W$^c$ & 
$\sim10^{7}$$^c$ & 
66\% \\

\makecell[c]{Hohn \textit{et al.}\cite{hohn2023energy}} & 
\makecell[c]{InGaAs QD} & 
\makecell[c]{XX-X cascade} & 
237 & 
$10^{4}$ & 
4.6 $\mu$W & 
$2.2 \times 10^{6}$ & 
73\% \\

\makecell[c]{Zhao \textit{et al.}\cite{zhao2020high}} & 
\makecell[c]{Thin-film \\ PPLN} & 
SPDC & 
$\sim$680 & 
$5.7 \times 10^{4}$ & 
$\sim$0.23 mW & 
$2.5 \times 10^{5}$ & 
\makecell[c]{99\% \\ (0.03 mW)$^d$} \\

\makecell[c]{Finco \textit{et al.}\cite{finco2024time}} & 
\makecell[c]{Thin-film \\ PPLN} & 
SPDC & 
-- & 
1800 & 
2 $\mu$W & 
$9 \times 10^{5}$ & 
78\% \\

\makecell[c]{Lu \textit{et al.}\cite{lu2019chip}} & 
Si$_3$N$_4$ & 
SFWM & 
\makecell[c]{26.8 \\ (CAR)} & 
$1.84 \times 10^{4}$ & 
$\sim$0.5 mW & 
$3.7 \times 10^{4}$ & 
83\% \\

\bottomrule
\end{tabular}

\vspace{4pt}
\footnotesize
\begin{minipage}{\textwidth}
\raggedright
$^a$ Detected pair rates are listed as measured, without correction for system efficiency. \\
$^b$ Data correspond to the central peak $|2\rangle$ only; inclusion of the $|1_t,1_{t+\tau_p}\rangle$ side peaks would yield higher values. \\
$^c$ Pump power not specified in the reference; values estimated from saturation parameters and given as order-of-magnitude approximations. \\
$^d$ The entanglement quality was characterized only at a pump power of 0.03 mW. When the pump power is increased, the probability of generating multiple photon pairs within the same time window inevitably rises, which in turn degrades the observed entanglement. Consequently, the Franson interference visibility is expected to decrease at the higher pump power of $\sim$0.23 mW used for bright-pair generation.\\
$^e$ Abbreviations:  Heitler, weak-excitation Heitler regime; F-P, Fabry--Pérot filter; XX-X, biexciton--exciton cascade; PPLN, periodically poled lithium niobate; SPDC, spontaneous parametric down-conversion; SFWM, spontaneous four-wave mixing; CAR, coincidence-to-accidental ratio; $\mathcal{V}$, two-photon interference visibility.
\end{minipage}
\end{table}

%%revised table

We summarize in Table \ref{tab:photon_pair_comparison} the state-of-the-art experimental platforms for generating time-energy entangled photon pairs. The key performance metrics include the photon-pair purity, the detected pair rate, the pump power, the detected pair rate per unit pump power, and the entanglement visibility. Here, the normalized brightness is defined as
$$
R=\frac{\mathcal{N}}{T P},
$$
where $\mathcal{N}$ is the number of detected photon pairs, $T$ is the integration time, and $P$ is the pump power. The value of $\mathcal{N}$ is extracted from the unnormalized second-order correlation histogram. As shown in Fig.~\ref{S7}, for an integration time of 300 s, the number of photon pairs is obtained by subtracting the background counts within 50 time bins from the total counts within the central 50 bins (bin width: 20 ps), yielding $\mathcal{N}=19998-177=19821$, which corresponds to a detected pair rate of 66.07 Hz.

\begin{figure}[htb]
\centering 
\includegraphics[width=0.55\textwidth]{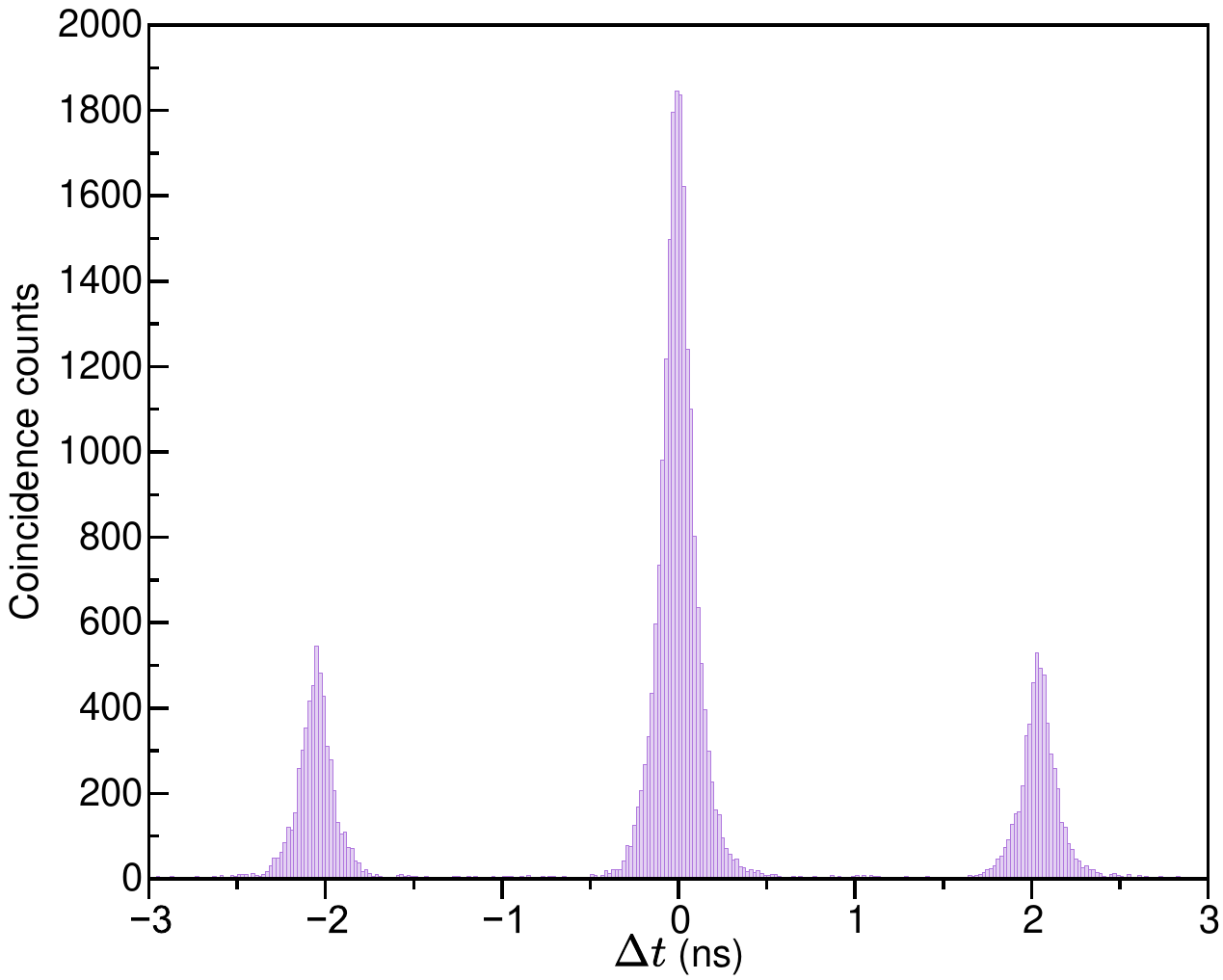}
\caption{\textbf{Raw autocorrelation data of the AMZI-prepared output, measured at $\bar{n}=0.01$.}}
\label{S7}
\end{figure}

Among the reported works, our scheme—based on quantum-dot RF in the Heitler regime combined with linear-optical interferometric state engineering using an AMZI—achieves a remarkable normalized brightness of $2.1\times10^9$~Hz/mW at an ultralow pump power of only 31.9 pW, with the AMZI-filtered state exhibiting an exceptionally high photon-number purity of $g^{(2)}_{HBT}(0)\approx511$. Compared to other nonlinear optical platforms, our system demonstrates a favorable combination of strong photon-number purity, high detected pair rate per pump power, and high Franson interference visibility $\mathcal{V}$, highlighting its potential for efficient and scalable photon-pair generation in quantum photonic applications.

 % 仅SI模式需要参考文献
    \ifonlySI
        \bibliography{mainref}
        \bibliographystyle{\natphysbibstyle}
    \fi
\fi

%%%%%%%%%%%%%%%% SUPPLEMENTARY TABLES %%%%%%%%%%%%%%%

%%%%%%%%%%% CAPTIONS FOR OTHER SUPPLEMENTARY FILES %%%%%%%%%%

%\clearpage % Clear all remaining figures and tables then start a new page

%%%%%%%%%%%%%%%% SUPPLEMENTARY REFERENCES %%%%%%%%%%%%%%%

% Do NOT include a reference list in the supplement.
% All references must be in a single list at the end of the main text.
% The copyeditors will ensure that the correct reference list appears with each version of the paper
% (print, HTML, PDF, mobile app, metadata for bibliographic databases etc.)

\end{document}